\documentclass[useAMS,usenatbib]{mn2e}
\usepackage{amsmath,amssymb,mathrsfs,graphicx,float,latexsym}
\usepackage{epstopdf,ragged2e}
\usepackage{hyperref}

\newcommand{\rs}{R_{\star}}

\newcommand{\ma}{M_{\text{a}}}
\newcommand{\Rm}{R_{\text{m}}}
\newcommand{\ri}{R_{\star}}
\newcommand{\rf}{R_{\text{f}}}

\newcommand{\bn}{b^{\text{N}}}
\newcommand{\bs}{b^{\text{S}}}
\newcommand{\man}{M_{\text{a}}^{\text{N}}}
\newcommand{\mas}{M_{\text{a}}^{\text{S}}}

\def\apj{{ApJ}}

\def\aap{{A\&A}}

\def\mnras{{MNRAS}}

\def\nat{{Nature}}

\def\prc{{Physical Review C}}

\def\04a{{2004 a}}
\def\04b{{2004 b}}

\title[Multipolar magnetospheres in recycled pulsars]{Recycled pulsars with multipolar magnetospheres from accretion-induced magnetic burial}
\author[A. G. Suvorov and A. Melatos]{
A. G. Suvorov$^{1}$\thanks{E-mail: arthur.suvorov@tat.uni-tuebingen.de} and
A. Melatos$^{2,3}$
\\
$^{1}$Theoretical Astrophysics, Eberhard Karls University of T{\"u}bingen, T{\"u}bingen, D-72076, Germany\\
$^{2}$School of Physics, University of Melbourne, Parkville, VIC 3010, Australia\\
$^{3}$Australian Research Council Centre of Excellence for Gravitational Wave Discovery (OzGrav), \\
\,University of Melbourne, Parkville, VIC 3010, Australia\\	
}
\begin{document}

\date{Accepted ?. Received ?; in original form ?}

\pagerange{\pageref{firstpage}--\pageref{lastpage}} \pubyear{?}

\maketitle\label{firstpage}

\begin{abstract}

\noindent{Many millisecond pulsars are thought to be old neutron stars spun up (`recycled') during an earlier accretion phase. They typically have relatively weak ($\lesssim 10^{9} \text{ G}$) dipole field strengths, consistent with accretion-induced magnetic burial. Recent data from the Neutron Star Interior Composition Explorer indicate that hot spots atop the recycled pulsar PSR J0030$+$0451 are not antipodal, so that the magnetic field cannot be that of a centered dipole. In this paper it is shown that multipolarity is naturally expected in the burial scenario because of equatorial field line compression. Grad-Shafranov equilibria are constructed to show how magnetic multipole moments can be calculated in terms of various properties, such as the amount of accreted mass and the crustal equation of state.}

\end{abstract}


\section{Introduction}

Advances in phase-resolved X-ray spectroscopy, primarily associated with the Neutron Star Interior Composition Explorer (NICER) \citep{ray19,riley19}, have allowed for surface temperature maps of rapidly rotating neutron stars to be measured with increasing accuracy. The maps reveal the location and geometry of `hot spots' with $\lesssim$ keV blackbody temperatures, which are produced when electrons (or positrons) created in charge-starved regions of the magnetosphere (`gaps') flow back along magnetic field lines and crash onto the stellar surface \citep{usov95,zhang97,hard02}. As a result the maps reveal detailed information about the magnetic field $\boldsymbol{B}$ on the surface and in the magnetosphere \citep{cheng03,miller19}; hot spot geometry can be tied to gap structure \citep{stur71,rud75,arons83,mus03}. For example, an analysis of thermal X-ray pulsations was performed by \cite{bilous19} using NICER data for the object PSR J0030$+$0451, a millisecond-pulsar `recycled' by accretion during an earlier low-mass X-ray binary (LMXB) stage \citep{alpar82,urpin98}. They found that an antipodal (i.e. diametrically opposed) hot spot model does not replicate the data, {which implies that particle backflow, and hence the magnetic field, cannot be reflection symmetric about the geographic equator. In particular, the magnetic field cannot be a centered dipole; it must include appreciable even-order multipoles.}

There are several physical mechanisms which may instigate magnetic multipolarity, such as crustal Hall drift \citep{gep13}, accelerated Ohmic diffusion \citep{urp95}, or accretion-induced magnetic burial \citep{bis74,blon86}. The last mechanism, the focus of this work, is a process where accreted material piles onto the polar cap, and gradually equilibrates toward a mountain-like mass distribution (`magnetic mountain') whose weight is supported by the compressed, equatorial magnetic field  \citep{brobil98,melphi01,pm04,zhang06,dip12}. The polar magnetic field is buried in the process, reducing the magnetic dipole moment, consistent with observations of LMXBs \citep{taam86,hart08,pat12,dip17}. Simultaneously the local magnetic field increases in strength and becomes multipolar. There is then an interesting question about whether the field stays buried or gradually resurrects, either rapidly due to hydromagnetic instabilities \citep{cumm01,litwin01,dip13} or slowly due to Ohmic \citep{konar04,vig09,wette10} or thermal \citep{brown98,suvmel19} relaxation. It is the purpose of this work, assuming that the field remains buried over sufficiently long time-scales\footnote{For isolated, recycled pulsars, such as PSR J0030$+$045, this amounts to assuming that the mountain remains, and that burial is sustained, after the companion has been ablated away by the relativistic pulsar wind \citep{ras89,bis90}.}, to demonstrate how the multipolarity of the magnetosphere may be tied to the accretion history of the pulsar, which depends on parameters such as the total accreted mass $M_{a}$ and the crustal equation of state (EOS) \citep{pri11}. 

Previous works have focussed mostly on the case of symmetric accretion. Recently, \cite{singh20} studied compositional gradients in asymmetrically accreted mountains [see also \cite{ush00,wij13,hask15}], and found that mass-density perturbations triggered by anisotropies in thermonuclear reaction rates can lead to levels of gravitational radiation which might account for the observed episodes of accelerated spindown in PSR J1023$+$0038 \citep{hask17}. However, \cite{singh20} considered low accreted masses ($10^{-15} \lesssim M_{a}/M_{\odot} \lesssim 10^{-11}$) where burial effects are negligible or just starting to make themselves felt. In particular, asymmetric flows depositing sufficiently large $M_{a}$ at the two poles would squash the equatorial band of the compressed $\boldsymbol{B}$ field away from the equator and make non-antipodal hot spots. We study this scenario in this paper. The EOS, and in particular the compressibility, of the accreted crust sets a characteristic height and mass density for the mountain, which naturally impact on the efficacy of burial \citep{pri11}. 

We model the magnetic field of isolated, recycled pulsars by computing hydromagnetic equilibria for a variety of magnetic mountain models using a version of the Grad-Shafranov solver initially developed by \cite{pm04}, which is extended to handle unequal mass fluxes in the northern and southern hemispheres. We consider neutron stars with pre-accretion magnetic fields that are dipolar for simplicity. The main new features of this work are that we consider north-south asymmetric accretion with masses $M_{a} \lesssim 10^{-4} M_{\odot}$, large enough to facilitate substantial burial, and show how to build self-consistent magnetohydrodynamic equilibria in this case. {This upper limit corresponds directly to the systems GX 1$+$4 and Her X$-$1, for example, which have $M_{a} \approx 10^{-4} M_{\odot}$ \citep{taam86}. However, our simulations may also reasonably describe higher-$M_{a}$ systems if some accreted material sinks into the star \citep{choud02,wette10}, so that only a fraction of the true $M_{a}$ is confined within the mountain (see Sec. 2.3).} We also show how one may relate the multipolarity of the buried magnetic field to astrophysical observables, relevant for existing and upcoming NICER X-ray timing data.


This short paper is organised as follows. In Section 2 we review the key features of magnetic burial, and describe the hydromagnetic structure of the permanent mountain that arises following an episode of accretion (Sec. 2.1), which depends on the mass-flux (Sec. 2.2) and crustal EOS (Sec. 2.3). The computational procedures involved and boundary conditions employed are detailed in Sec. 2.4. Global multipole moments (Section 3) are calculated for the symmetric (asymmetric) case in Section 4 (Section 5) for various EOS and accreted masses. Preliminary interpretations of the recent NICER results are given in Section 6.

\section{Magnetic burial}

During accretion, the neutron star's magnetic field lines are compressed towards the equator by infalling matter. As a result the post-accretion magnetic field is more intense than the natal field locally at the equator, but the dipole moment (a global quantity) is reduced because it is dominated by the polar magnetic field, which is `buried' by accretion. An inverse relationship between the accreted mass $M_{a}$ and the global dipole moment $\mu_{1}$ has been observed in LMXBs \citep{taam86,hart08,pat12}, qualitatively consistent with numerical magnetohydrodynamics simulations of magnetic burial \citep{pm04,vig09,pri11,dip12,dip17}. The exact relationship between $M_{a}$ and the resulting magnetic field geometry depends on the hydromagnetic structure of the mountain, which we explore in Sec. 2.1.

\subsection{Hydromagnetic structure}

The mountain equilibrium is determined by the force balance (Grad-Shafranov) equation 
\begin{equation} \label{eq:gradshaf}
0 = \nabla p + \rho \nabla \Phi + \Delta^2 \psi \nabla \psi,
\end{equation}
where $\Delta^2 = \left(4 \pi r^2 \sin^2\theta \right)^{-1} \left[ \partial_{rr} + r^{-2}  \sin \theta \partial_{\theta} \left( \csc\theta \partial_{\theta} \right) \right]$ is the Grad-Shafranov operator in spherical coordinates $(r,\theta,\phi)$, $p$ is the fluid pressure, $\rho$ is the mass density, and $\psi$ is a scalar flux defining the magnetic field,
\begin{equation} \label{eq:magfield}
\boldsymbol{B} = \frac{ \nabla \psi \times \hat{\boldsymbol{e}}_{\phi} } {r \sin \theta}.
\end{equation}
As in previous studies \citep{pri11,suvmel19,singh20}, we prescribe a spherically symmetric gravitational potential with constant acceleration $\Phi \propto r$, and ignore the self-gravity of the mountain (i.e., we do not solve the Poisson equation) for simplicity \citep{haskell06,yoshida13}. Solutions to equation \eqref{eq:gradshaf} self-consistently determine the hydrostatic pressure of the mountain, which balances the Lorentz force coming from the twisted magnetic field, when an EOS (Sec. 2.3) and appropriate boundary conditions (Sec. 2.4) are imposed. 

In general, \eqref{eq:gradshaf} can be solved exactly to give
\begin{equation} \label{eq:lagcha}
\int \frac {d p} {d \rho} \frac {d \rho} {\rho} = F(\psi) - \left( \Phi - \Phi_{0} \right),
\end{equation}
where $\Phi_{0} = \Phi(R_{\star})$ is a reference potential at stellar radius $R_{\star}$, and $F$ is a function of the scalar flux set by the accretion physics. In particular, the pre- and post-accretion states can be related through the flux-freezing constraint that the mass-flux ratio $d M / d \psi$, defined as the mass confined within the volume between infinitesimally separated flux surfaces $\psi$ and $\psi + d \psi$, matches that of the pre-accretion state plus any accreted matter \citep{mous74,melphi01}, which leads to
\begin{equation} \label{eq:fluxfreezing}
\frac {d M} {d \psi} = 2 \pi \int_{C} ds \rho \left[ r(s), \theta(s) \right] r \sin \theta | \nabla \psi |^{-1} ,
\end{equation}
where $C$ is the curve $\psi[r(s),\theta(s)] = \psi$ parametrised by the arc length $s$. Equation \eqref{eq:fluxfreezing} can be solved by inverting \eqref{eq:lagcha} for $\rho$ in terms of $\psi$ given a barotropic relation $p=p(\rho)$, leading to a unique expression for the function $F(\psi)$ given $M(\psi)$; see \cite{pri11} for more details. 

\subsection{Mass-flux ratio with and without north-south symmetry}

The exact form for the mass-flux $M(\psi)$ is determined by the details of how plasma flows from the accretion disc through the magnetosphere \citep{gosh79}, which depends on a number of largely unknown factors, such as the onset of magnetic turbulence \citep{bal91,abar18} {or other magnetohydrodynamic instabilities \citep{lasota01}, the magnetization parameter of bulk motion in the pulsar wind \citep{ang10,kong11} (see also Sec. 3), and the inclination angle between the star's magnetic axis and the disc's rotation axis \citep{roman03} [see \cite{don07} for a review]}. We adopt the approximation that most of the accreted material is distributed nearly uniformly within the polar-flux tube $0 \leq \psi \leq \psi_{a}$, where $\psi_{a}$ labels the last field line that closes inside the inner edge of the accretion disc. The mass-flux distribution in one hemisphere is set, as in previous works \citep{pm04,mp05,suvmel19}, as
\begin{equation} \label{eq:massfluxexp}
M(\psi) = \frac {M_{a} \left( 1 - e^{-\psi/\psi_{a}}\right)} {2 \left( 1 - e^{-b}\right)},
\end{equation}
with $b = \psi_{\star} / \psi_{a}$ for $\psi_{\star} = B_{\star} R_{\star}^2 /2$ given a natal polar field strength $B_{\star}$. 

In the case of symmetric accretion, the profile \eqref{eq:massfluxexp} with the same values of $M_{a}$ and $b$ is adopted in both hemispheres, i.e. we set $\bn = \bs$ and $\man = \mas$, where the superscripts N and S denote north and south, respectively. If $b=\bn$ in the northern hemisphere is higher (say) than $b=\bs$ in the southern hemisphere, a lower percentage of $\man$ accretes within the northern polar cap $0 \leq \psi(r,\theta \leq \pi/2) \leq \psi_{a}^{\text{N}}$ than $\mas$ accretes within the southern polar cap. More explicitly, the form \eqref{eq:massfluxexp} for $M(\psi)$ implies that the amount of accreted mass within one flux tube $0 \leq \psi \leq \psi_{a}^{\text{N,S}}$ reads
\begin{equation} \label{eq:bformula}
\int^{\psi_{a}^{\text{N,S}}}_{0} \frac{ d M} {d \psi} d \psi = \frac {\left(e -1 \right) e^{b^{\text{N,S}}-1} } { e^{b^{\text{N,S}}} - 1} M_{a}^{\text{N,S}}.
\end{equation}
From \eqref{eq:bformula} we see that all of $M_{a}^{\text{N,S}}$ accumulates within the cap for $b^{\text{N,S}} =1$, while $\approx 0.63 M_{a}^{\text{N,S}}$ accumulates there for $b \gg 1$.

It is important to note that there is one polar-flux tube per hemisphere, and that $\psi$ is reflection symmetric (antisymmetric) about the geographic equator for a pure odd- (even$-)$order multipole. A north-south asymmetry in the form of different $b$ and $M_{a}$ values could naturally arise if, for example, accretion episodes are sporadic and the disc is tilted at some angle relative to the star \citep{str96,roman03}, resulting in an asymmetric accretion flow.  Because the reflection symmetry about the equator of $\psi$ is broken under these circumstances, non-antipodal hot spot structures can arise naturally. The details of the numerical solution of \eqref{eq:gradshaf} and \eqref{eq:fluxfreezing} in this case are given in Sec. 2.4.

\subsection{Crustal equation of state}

We adopt a polytropic EOS, $p = K \rho^{\Gamma}$, where $K$ is measured in cgs units (dyn $\text{g}^{-\Gamma}$ $\text{cm}^{3 \Gamma -2 }$) and $\Gamma$ is the adiabatic constant which is tied to the accretion history in the following sense. For low accretion rates $\dot{M}_{a} \lesssim 10^{-10} M_{\odot} \text{ yr}^{-1}$, accreted plasma is approximately isothermal $(\Gamma \approx 1)$ as the crustal temperature is locked to that of the core \citep{miralda90}. At near-Eddington rates $\dot{M}_{a} \gtrsim 10^{-8} M_{\odot} \text{ yr}^{-1}$, neutrino cooling, bremsstrahlung, and electron capture by nuclei generate thermal gradients throughout the crust and lead to density-dependent $K$ and $\Gamma$ factors \citep{brown98,brown00,meisel18}. In an actively accreting system, the observed X-ray flux $F_{X} \approx G M \dot{M}_{a} / \left( 4 \pi R_{\star}  d^2 \right)$, where $d$ is the distance to the star, can be used to estimate $\Gamma$ and thus the EOS \citep{pot19}. However, for recycled pulsars having concluded their accretion episodes, modelling the crustal EOS as a polytrope with a constant value of $\Gamma$ is likely a fair approximation \citep{haen90,brown98,haen18}. 

Calculations by \cite{haen90} using a compressible liquid drop model \citep{mackie77} to estimate thermodynamic rates\footnote{See also \cite{haen18} for recent results using an energy-density functional approach.} show that the EOS is dominated by the contribution from relativistic, degenerate electrons at sub-neutron-drip [$\rho_{\text{nd}} \sim 5 \times 10^{11} \text{ g cm}^{-3}$ \citep{cham15}] densities. Mountains with maximum densities below this threshold can be described as an isentropic gas of electrons, with polytropic EOS $p = K \rho^{\Gamma}$ for $K = 4.93 \times 10^{14}$ (henceforth, cgs units for this quantity are always assumed) and $\Gamma = 4/3$. We find \emph{a posteriori} that $\rho \lesssim \rho_{\text{nd}}$ corresponds to accreted masses $M_{a} \lesssim 2 \times 10^{-7} M_{\odot}$. It is important to note that, in reality, the total amount of mass accreted during an LMXB phase is likely much higher than the above $M_{a}$ value, though most of the material would gradually sink into the lower-density substrate \citep{choud02,wette10}. Sinking is not modelled here as it is beyond the scope of this work and requires detailed assumptions about the layered structure of the neutron star. Strictly speaking, therefore, $M_{a}$ should be thought of as the amount of accreted mass that remains within the crust after sinking occurs.

For accreted masses $M_{a} \gg 10^{-7} M_{\odot}$, with $\rho \gtrsim \rho_{\text{nd}}$ at the base of the mountain, a more realistic EOS can be constructed by applying a damped least-squares algorithm to the data collated in Table 1 of \cite{haen90} to determine best-fit values for $K$ and $\Gamma$. This was done by \cite{suvmel19}, who found $\Gamma = 1.18$ and $K = 6.18 \times 10^{14}$. For these choices, the relative differences between the pressures for the polytropic EOS and the \cite{haen90} values for $\rho \gtrsim 10^{12} \text{ g cm}^{-3}$ are $\lesssim 10\%$, with decreasing errors for increasing $\rho$; see Fig. 2 in \cite{suvmel19}. This EOS is referred to as the ``accreted crust EOS'' throughout and applies for accreted masses $M_{a} \gtrsim 10^{-5} M_{\odot}$.

\subsection{Boundary conditions with and without north-south symmetry}

When accretion occurs symmetrically in the northern and southern hemispheres, the boundary conditions we adopt are, as in previous works \citep{pm04,pri11,suvmel19}, $\psi(\ri,\theta) = \psi_{\star} \sin^2 \theta$ (surface dipole condition), $\partial \psi / \partial r (\rf,\theta) = 0$ (outflow), $\psi(r,0) = 0$ (straight polar field line), and $\partial \psi/ \partial \theta (r, \pi/2) = 0$ (north-south symmetry), where $\ri \leq r \leq \rf$ and $0 \leq \theta \leq \pi/2$ delimit the computational grid. We do not simulate the region $\pi/2 \leq \theta \leq \pi$ because the Neumann condition $\partial \psi/ \partial \theta (r, \pi/2) = 0$ forces $\psi(r,\theta) = \psi(r,\pi-\theta)$. The maximum radius $\rf$ is chosen large enough to encompass most of the diamagnetic screening currents (typically $\rf \approx 1.3 R_{\star}$) \citep{pm04,mast12}, so that we can continuously match the $\psi$ profile outside of the computational region to a sum of force-free multipoles. In any case, we find that changing the value of $\rf$ by $\lesssim 50 \%$ has little effect on the results: quantitative aspects of the multipolarity are approximately unchanged (see Secs. 3 and 5.1).

North-south symmetry necessarily precludes the existence of even-order multipoles, which implies north-south symmetric backflow heating and antipodal hot spots. By contrast, thermal X-ray pulsations from PSR J0030$+$0451 show evidence for non-antipodal hot spots \citep{bilous19} [see also \cite{cheng03}]. To match the observations qualitatively, we need to include north-south asymmetry in the model. One possible approach is to prescribe a different $M(\psi)$ in the two hemispheres provided that care is taken to preserve continuity across the equator. Attempting naively to run separate north and south simulations with different $b$ or $M_{a}$ values, to build a `total' solution which has a different flux profile $M(\psi)$ in the north and south (i.e. `gluing' two profiles together), leads to an unphysical $\psi$ which is discontinuous at the equator.

To overcome this, we first run a simulation in the northern (or southern, equivalently) hemisphere alone with the symmetric boundary conditions listed above. The solver returns a flux profile at the equator $\psi(r,\pi/2) = \xi(r)$ (say). Re-running the same single-hemisphere simulation but with the boundary condition $\psi(r,\pi/2) = \xi(r)$ instead of the north-south Neumann condition returns the same solution -- as a numerical check -- within an accuracy of $\lesssim 2\%$. The small discrepancy arises because we fit a cubic-spline to the output of the initial simulation to construct $\xi(r)$. A full simulation over both hemispheres with $\bn \neq \bs$ and/or $\man \neq \mas$ is then run with the straight polar field line condition $\psi(r,\pi) = \psi(r,0)$, and the function $\xi(r)$ is used as an internal condition to force $\psi(r,\pi/2) = \xi(r)$. This ensures that we construct a solution over $0 \leq \theta \leq \pi$ that is everywhere continuous and conforms to the single-hemisphere simulations in the symmetric limit. 

\begin{figure*}
\includegraphics[width=\textwidth]{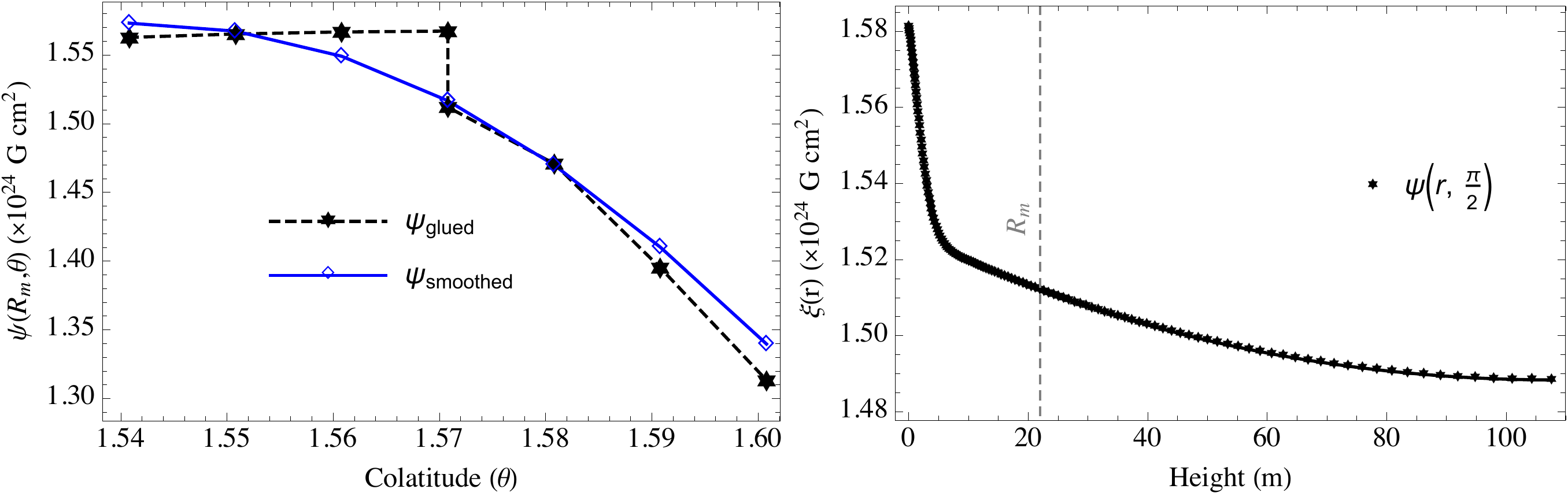}
\caption{Comparison of a naively `glued' flux profile $\psi(\Rm,\theta)$ and a smoothed one (left panel) in the vicinity of the equator $\theta \approx \pi/2$ obtained from a simulation employing the equatorial Dirichlet condition $\psi(r,\pi/2) = \xi(r)$ (right panel) for a representative example of asymmetric accretion; see text for details on simulation parameters and the `gluing' and `smoothing' procedures. The maximum mountain height $\Rm$ is shown by the dashed, gray line.}
\label{fig:psifit}
\end{figure*}

The left panel of Figure \ref{fig:psifit} contrasts two $\psi$ profiles, one obtained by `gluing' naively (black stars) as described above and a second `smoothed' profile (blue diamonds) obtained using the periodic boundary condition together with the internal flag $\psi(r,\pi/2) = \xi(r)$. The cubic-spline $\xi(r)$ for these simulations (though not used in the `naive' case) is shown in the right panel as the solid curve, which interpolates between grid points (black stars). For both the `glued' and `smoothed' cases, the function $\psi$ is evaluated at the mountain-atmosphere boundary $\Rm$, defined as the first radial grid point where $\rho$ vanishes (to numerical precision) for all $\theta$. For this simulation we set $\man = 10^{-5} M_{\odot}$, $\bn=3$ and $\mas = 2 \times 10^{-5} M_{\odot}$, $\bs=4$, and use the accreted crust EOS described in Sec 2.3 (see Sec. 5.1 for further properties pertaining to this particular solution). For these parameters we find $\Rm = 22$ m, shown by the dashed line in the right-hand panel. The maximum disagreements between $\psi$ profiles are of the order $\lesssim 2\%$ at $r = \Rm$, dropping to $\lesssim 1\%$ at smaller radii, and the continuity of $\psi$ at $\theta = \pi/2$ can be clearly seen for the `smoothed' run adopting the periodic conditions instead of the unphysical Neumann condition. 

Using the methods detailed in this section, we are now in a position to construct equilibrium profiles for magnetic mountains. This is achieved by extending the Grad-Shafranov solver initially developed by \cite{pm04} and \cite{pri11} to incorporate the periodic [$\psi(r,0) = \psi(r,\pi)$] and internal [$\psi(r,\pi/2) = \xi(r)$] conditions. For all simulations presented here, we employ a static, logarithmic radial grid with $N_{r} = 256$ cells, which provides sufficient resolution to capture flux gradients in the base of the mountain, where the mass density is greatest. A uniform, polar grid in $\cos \theta$ with $N_{\theta} = 512$ points is used, with $N_{\theta}/2$ cells in each hemisphere, which adequately captures the lateral pressure gradients and allows us to check continuity across the equator, as in Fig. \ref{fig:psifit}. {The exact convergence criteria, and their relation to the numerical mesh size, are laid out in Appendix A of \cite{pri11} and are satisfied for all simulations performed here with a $256 \times 512$ grid. In several cases where an analytic solution to the Grad-Shafranov problem is available, \cite{pm04} found that a resolution of $128 \times 256$ is already sufficient to reduce the mean errors in $\psi$ to be less than $0.1\%$ at each cell; see Table B2 therein.} For further details about the explicit integration schemes used to solve \eqref{eq:gradshaf} while preserving flux-freezing \eqref{eq:fluxfreezing}, the reader is referred to \cite{suvmel19}. 

\section{Magnetic multipoles}

In this paper, we are interested in field multipolarity in recycled pulsars caused by accretion-induced magnetic burial, with the ultimate goal of interpreting NICER hot spot data. Given a solution $\psi$ to the Grad-Shafranov problem \eqref{eq:gradshaf} from some physical input (e.g., $M_{a}$, $b$, $B_{\star}$, $\Gamma$, $K$) we can match the solution to a force-free sum of multipoles at the outer boundary of the computational box, $r=\rf$. Recalling that $\rf$ is chosen to be sufficiently large so as to encompass most of the diamagnetic screening currents \citep{pm04,mast12}, the matching procedure provides a fair approximation to the external, vacuum field geometry post-accretion. Magnetospheric factors, such as the static \cite{gj69} plasma density or conduction currents \citep{arons83} near the light-cylinder prevent the external region from being vacuum in reality, though a modelling of these factors is beyond the scope of this work; see \cite{vig15} for a detailed discussion of gap physics and related factors. 

\begin{figure}
\includegraphics[width=0.497\textwidth]{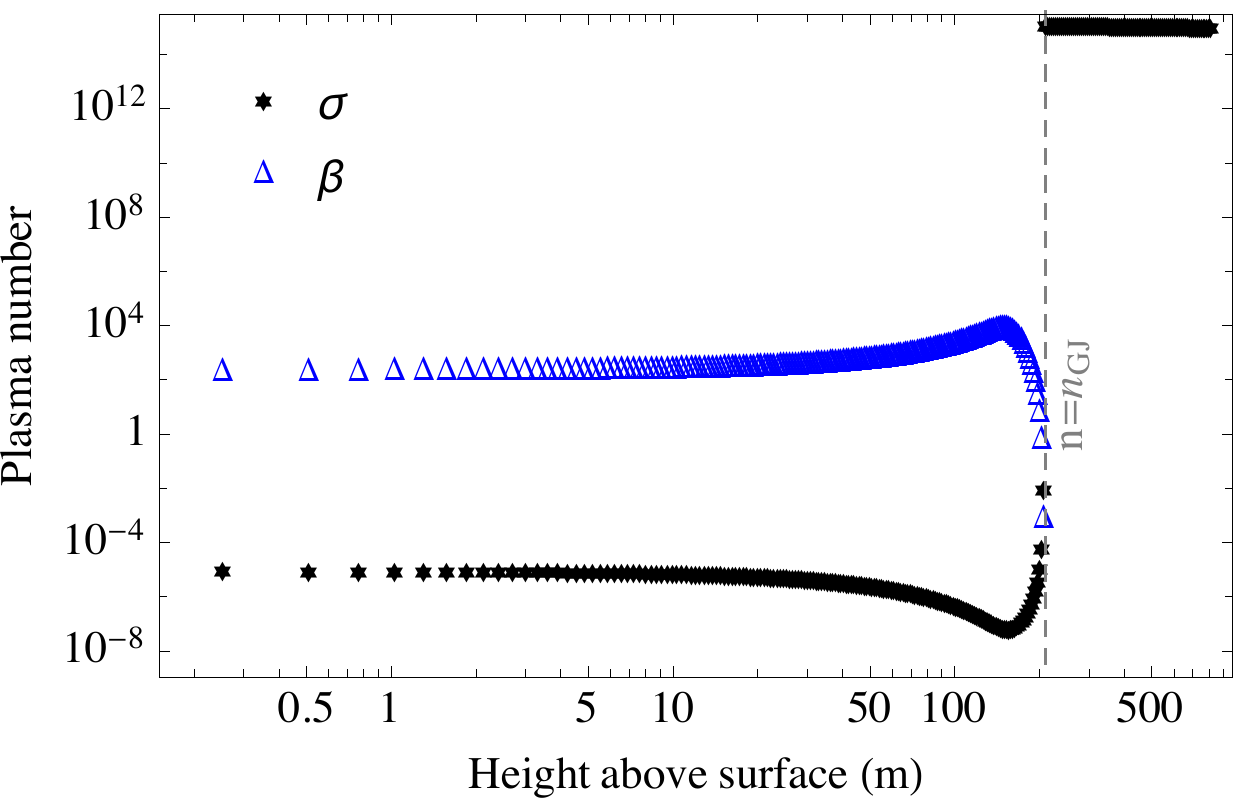}
\caption{Dimensionless plasma numbers $\sigma$ (black stars) and $\beta$ (blue diamonds) at each radial grid point for $\theta = 0.1$ for the relativistic, electron EOS with $M_{a} = 1.2 \times 10^{-7} M_{\odot}$. The mountain-atmosphere boundary, where the particle number density $n$ drops to the Goldreich-Julian value $n_{\text{GJ}}$, is shown by the dashed line.}
\label{fig:plasmanumbers}
\end{figure}

{We can nevertheless estimate the validity of the force-free-atmosphere assumption by tracking the value of the \emph{magnetization parameter} $\sigma$, defined as the ratio of the magnetic enthalpy density to the thermal enthalpy density. A large value of $\sigma$ relative to the bulk Lorentz factor indicates a force-free plasma \citep{michel91}. In general, we have $\sigma = B^2 / \left( 4 \pi h m_{e} n c^2 \right)$ for electron mass $m_{e}$, speed of light $c$, specific enthalpy $h$, and particle number density $n$. Another related quantity is the \emph{plasma beta}, $\beta$, defined as the ratio of hydrostatic pressure to magnetic pressure. Figure \ref{fig:plasmanumbers} shows $\sigma$ and $\beta$ for spin frequency $\nu = 205 \text{ Hz}$, as appropriate for PSR J0030$+$0451 \citep{riley19}, at each radial grid point near the northern magnetic pole $(\theta = 0.1)$ for a mountain with the relativistic electron EOS and $M_{a} = 1.2 \times 10^{-7} M_{\odot}$. To prevent discontinuities in the plasma numbers, a tenuous atmosphere has been included by setting a mass density floor equal to the Goldreich-Julian value. For $r - R_{\star} \gtrsim 208 \text{ m}$, where $n$ falls below the Goldreich-Julian value for this particular latitude, the magnetization is very large relative to unity $(\sigma \gtrsim 10^{13})$. In contrast, the plasma beta varies substantially throughout the bulk of the mountain $(10^{-2} \lesssim \beta \lesssim 10^{2})$, indicative of a complicated hydromagnetic structure, decreasing with altitude as expected. Figure \ref{fig:plasmanumbers} is typical for simulations performed in this work. }

We are now in a position to compute the moment structure. The $\ell$-th magnetic multipole moment is defined through
\begin{equation} \label{eq:multmoms}
\hspace{-0.2cm}\mu_{\ell}(r)= \frac {\ell \left( 2 \ell + 1 \right) r^{\ell}} {2 \left( \ell +1 \right)} \int^{1}_{-1} d \left( \cos \theta \right) \psi \left( r, \cos \theta \right) \frac {d P_{\ell}\left(\cos \theta \right)} {d \left( \cos \theta \right)},
\end{equation}
for Legendre polynomials $P_{\ell}$. The pre-accretion field, modelled as a pure dipole, has only $\mu_{1}(r \geq R_{\star}) = \mu_{\star} = B_{\star} R_{\star}^3/2$ non-zero. Here and throughout we take the natal polar field strength $B_{\star} = 10^{12.5}$G in line with population synthesis models [see, e.g., \cite{kaspi08}]. For presentation purposes, it is convenient to introduce normalised moments $\tilde{\mu}_{\ell}$ defined through
\begin{equation} \label{eq:normmom}
\tilde{\mu}_{\ell} = \frac{ 2 \left( \ell + 1 \right)} { \rf^{\ell - 1} \ell \left( 2 \ell + 1 \right)} \mu_{\ell}(\rf) ,
\end{equation}
so that each $\tilde{\mu}_{\ell}$ has the same units ($\text{G cm}^{3}$), and does not appear artificially inflated at large $\ell$ due to the $\ell$-dependent prefactor used in the standard definition \eqref{eq:multmoms}.

We compute a variety of magnetic equilibria assuming north-south symmetry for various accreted masses (Sec. 4.2) and EOS parameters (Sec. 4.3), and determine the moments $\tilde{\mu}_{\ell}$ of the resulting solutions using expression \eqref{eq:normmom}. Non-antipodal models without north-south symmetry are then presented in Sec. 5 along the same lines using the methods described in Sec 2.4. Given an angular velocity vector $\boldsymbol{\Omega}$ as well, one can determine gap structures from the resulting $\psi$ in principle, to eventually build a model of hot spot geometry from astrophysical observables. 

\section{Symmetric accretion: antipodal hot spots}

In this section we consider mountains built with north-south symmetric mass-flux $M(\psi)$, i.e. antipodal models. Although not directly applicable to the NICER results for PSR J0030$+$0451, these models allow us to investigate the impact of the crustal EOS and $M_{a}$ on the multipolarity, and provide reference values for the non-antipodal cases given in Sec. 5.

\subsection{Representative example}

We first consider a representative example to demonstrate the qualitative features of the multipolarity induced by magnetic burial. For this case, we take the relativistic degenerate electron EOS ($\Gamma = 4/3, K = 4.93 \times 10^{14}$) with $\bn = \bs = 3$ and $\man = \mas = 0.8 \times 10^{-7} M_{\odot}$, so that the total accreted mass is given by $\ma = 1.6 \times 10^{-7} M_{\odot}$.

\begin{figure*}
\includegraphics[width=\textwidth]{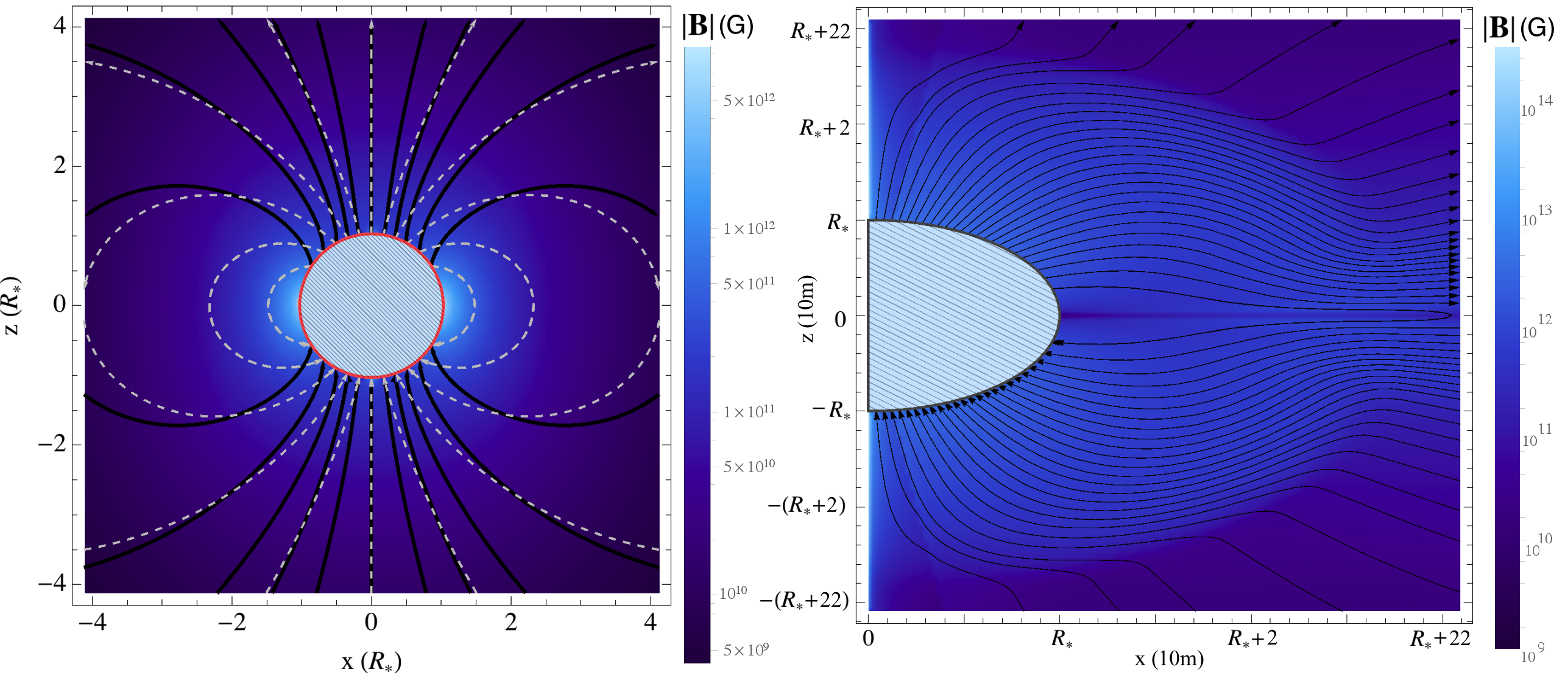}
\caption{Field lines in a meridional cross section for the pre- (dashed) and post- (solid) accretion magnetic fields for the symmetric case with $M_{a} = 1.6 \times 10^{-7} M_{\odot}$ and a crust with a relativistic electron EOS (i.e. $\Gamma = 4/3, K = 4.93 \times 10^{14}$). The colour scales show $|\boldsymbol{B}|$ for the post-accretion field \eqref{eq:magfield}, where brighter shades indicate a greater field strength. The neutron star surface is shown by the grayed-out surface centered at the origin. Field lines in the mountain region $\rs \leq r \leq \Rm$ (for $x>0$; right panel) are contained within the thin, red layer surrounding the star (left panel). The mountain achieves a maximum radial extent of $\Rm = 217 \text{ m}$ in this case.}
\label{fig:symb}
\end{figure*}

The left panel of Figure \ref{fig:symb} shows magnetic field lines (solid curves) in a meridional cross-section for a force-free sum of multipoles (up to order $\ell = 51$) fitted to the Grad-Shafranov output $\psi$ at the computational edge $\rf$. We take $\rf = 1.3 \rs$ so that the Grad-Shafranov simulation encompasses most of the diamagnetic screening currents. The outer edge $\rf$ exceeds the maximum radial extent of the mountain, $\Rm$, defined as the first $r$ such that $\rho$ vanishes for all $\theta$. For this simulation, we find $\Rm = 217 \text{ m}$. The mountain occupies the thin, red region surrounding the star, itself represented by the grayed-out surface centered at the origin. For comparison, field lines for the (dipolar) pre-burial field with the same crustal footpoints are shown by the dashed, gray lines, while the colour scale shows $|\boldsymbol{B}|$ of the post-burial field \eqref{eq:magfield}. Magnetic field lines in the mountain region $\rs \leq r \leq \Rm$ are shown in the right panel of Fig. \ref{fig:symb}. The field there reaches a maximum strength of $|\boldsymbol{B}| = 2.7 \times 10^{14} \text{ G}$, exceeding the field in the vacuum region $r \geq \Rm$ by almost two orders of magnitude. Field lines in the equatorial zone are pinched together, so that the overall geometry resembles a `magnetic tutu' configuration, familiar from previous simulations of magnetic burial \citep{melphi01,pm04}.

\begin{figure}
\includegraphics[width=0.497\textwidth]{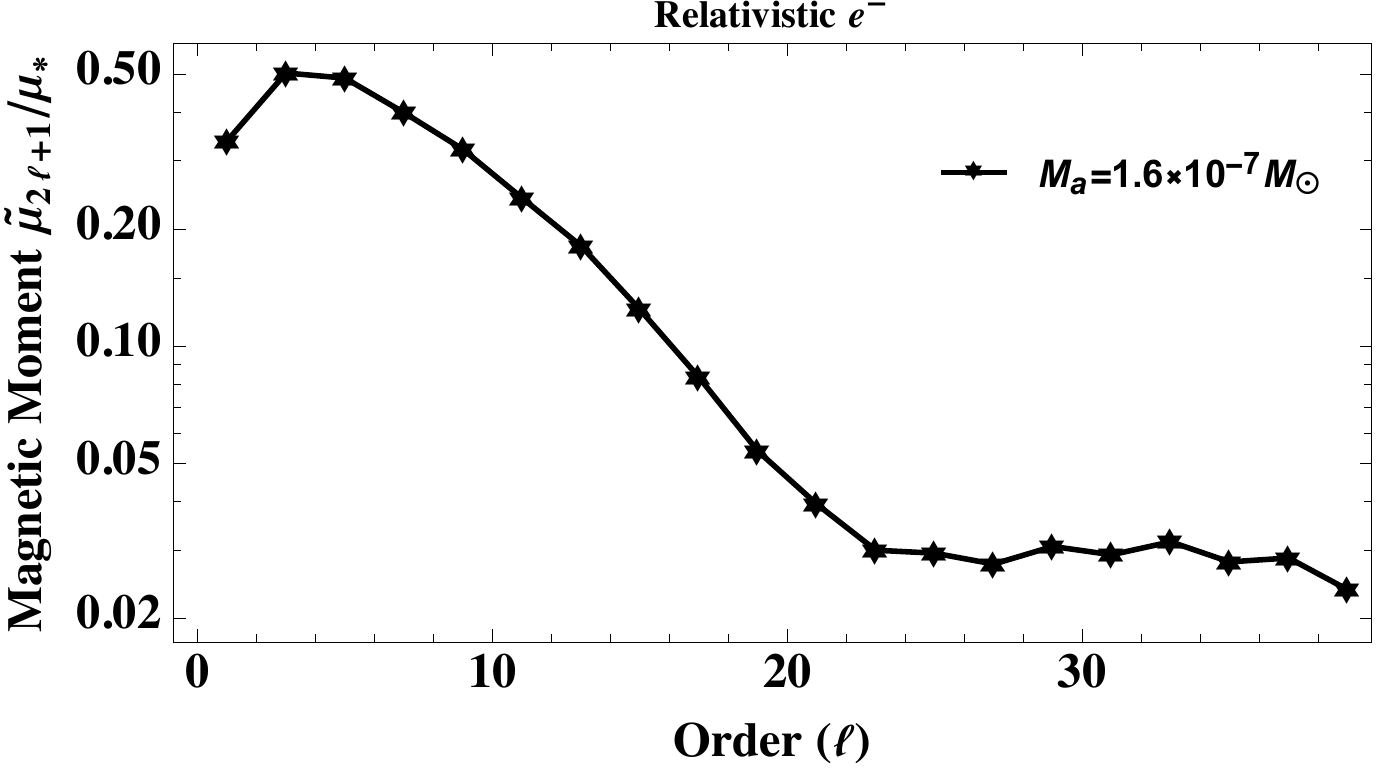}
\caption{Dimensionless multipole moments $\tilde{\mu}_{\ell}/\mu_{\star}$ versus multipole order $\ell$, computed from expression \eqref{eq:normmom}, for the post-accretion field displayed in Fig. \ref{fig:symb}.}
\label{fig:repe}
\end{figure}

The impact of burial on the global, vacuum field is most noticeable through the curvature of field lines with crustal footpoints near the equator. Field lines are combed sideways in the mountain region (right panel) due to accretion, which forces the matched field lines (left panel) to be stretched around the equator. In particular, only one of the field lines with this specific set of crustal footpoints closes within the radius $r \lesssim 4 R_{\star}$, while three such lines close for the pre-burial field. Figure \ref{fig:repe} shows the normalised multipoles $\tilde{\mu}_{\ell}$ up to $\ell = 39$ for this particular configuration, where we note again that only odd multipoles are non-zero for a north-south symmetric solution. The global dipole moment is buried by $67\%$ relative to the initial moment $\mu_{\star}$ (i.e. $\tilde{\mu}_{1} = 0.33 \mu_{\star}$), and the $\ell=3$ moment with $\tilde{\mu}_{3}/\mu_{\star} = 0.5$ is greatest, though the $\ell =5$ and $\ell=7$ moments with $\tilde{\mu}_{5}/ \mu_{\star} = 0.49$ and $\tilde{\mu}_{7}/\mu_{\star} = 0.40$, respectively, are also comparable to the octupole and exceed the dipole moment. Moments with $\ell \geq 11$ are weaker than $\tilde{\mu}_{1}$, and we find that $\tilde{\mu}_{\ell}$ decreases monotonically with $\ell$ for $5 \leq \ell \lesssim 25$, oscillates around $\sim 0.03 \mu_{\star}$ for $25 \lesssim \ell \leq 39$, and finally vanishes for $\ell > 39$. 

It is worth pointing out that, despite the field line broadening effect (i.e. fewer field lines appear to close for the post-accretion field) seen in Fig. \ref{fig:symb}, the higher-multipoles contribute negligibly to the stellar spindown except for very high rotational velocities $\Omega$. In general, a pure $\ell$-pole leads to a spin-down rate $\dot{\Omega} \sim \dot{\Omega}_{\text{dipole}} \left( R_{\star} \Omega / c \right)^{2 \ell - 2}$ \citep{arons93}, where $\dot{\Omega}_{\text{dipole}}$ is the dipole spin down rate. Even for the millisecond object PSR J0030$+$0451, the term within the parenthesis is of the order $\sim \left( 5.5 \pm 0.5 \right) \times 10^{-2}$ \citep{riley19}, so that $\dot{\Omega}_{\text{dipole}} \propto \mu_{1}^2$ dominates. 

The buried dipole moments $\mu_{1}$ we find for this and subsequent simulations are consistent with those reported in previous studies \citep{pm04,pri11,suvmel19}, and qualitatively agree with observations. Spindown measurements indicate that the magnetic dipole moment $\mu_{1}$ is inversely proportional to the accreted mass $M_{a}$ \citep{taam86,hart08,pat12}, in line with the empirical law proposed by \cite{shibaz}, $\mu_{1} = \mu_{\star} \left( 1 + M_{a} / M_{c} \right)^{-1}$, where $M_{c}$ is some characteristic mass which depends on the EOS. For the relativistic electron EOS, we have $M_{c} \sim 10^{-7} M_{\odot}$ \citep{pri11}.

\subsection{Accreted mass}

Here we consider a variety of models for fixed EOS parameters but for varying accreted masses, and compute the multipolarities of the exterior, force-free solutions fitted to Grad-Shafranov equilibria. For runs performed in this and the following section, we take $\man = \mas$ and $\bn = \bs = 3$.

\begin{figure}
\includegraphics[width=0.473\textwidth]{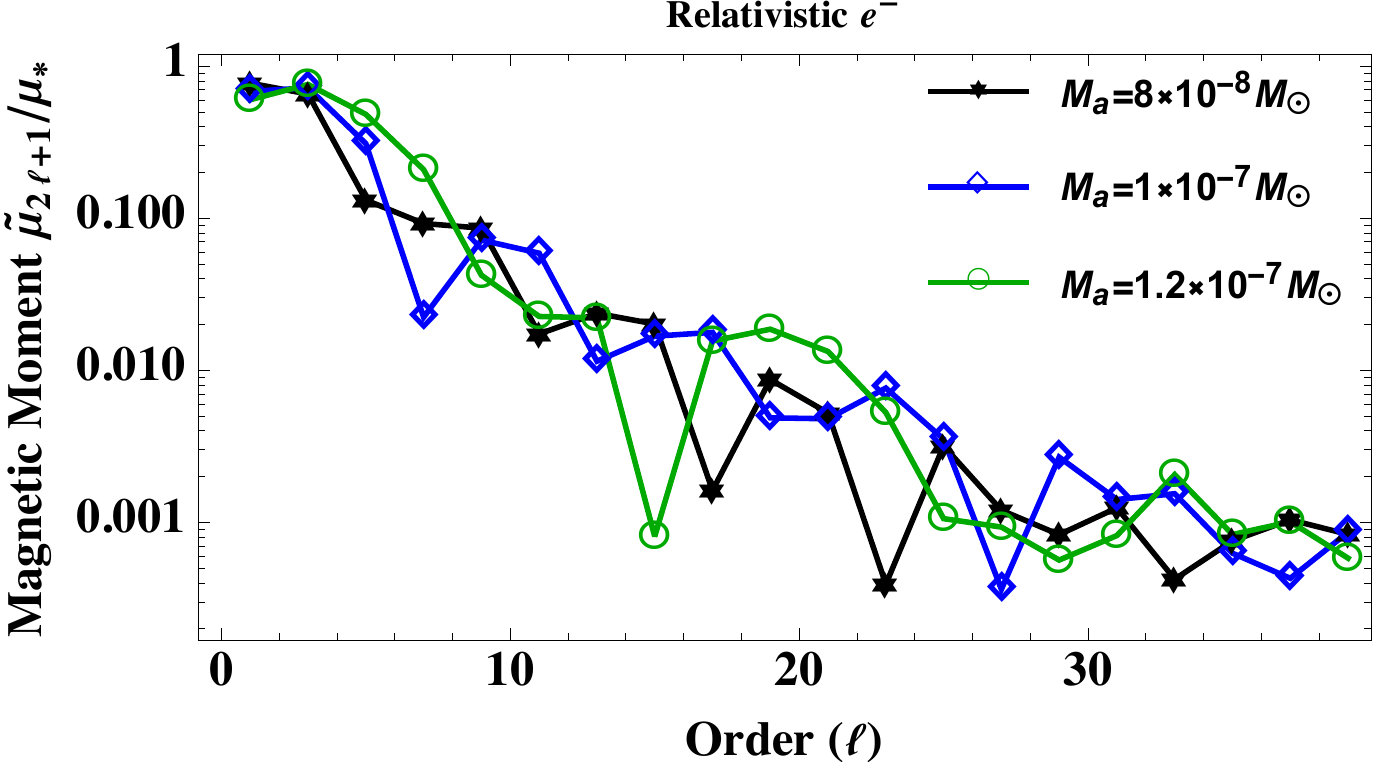}
\caption{Dimensionless multipole moments $\tilde{\mu}_{\ell}/\mu_{\star}$ versus multipole order $\ell$ for $M_{a} = 8 \times 10^{-8} M_{\odot}$ (black stars), $M_{a} = 10^{-7} M_{\odot}$ (blue diamonds), and $M_{a} = 1.2 \times 10^{-7} M_{\odot}$ (green circles) for a crust with polytropic EOS with adiabatic index $\Gamma = 4/3$ and polytropic constant $K = 4.93 \times 10^{14}$, as appropriate for an isentropic gas of relativistic electrons. \label{fig:symevarymass}
}
\end{figure}

Figure \ref{fig:symevarymass} shows the first $\ell \leq 39$ magnetic moments $\tilde{\mu}_{\ell}$ for the same EOS used in the representative example (cf. Fig. \ref{fig:repe}) but for $M_{a} = 8 \times 10^{-8} M_{\odot}$ (black stars), $M_{a} = 10^{-7} M_{\odot}$ (blue diamonds), and $M_{a} = 1.2 \times 10^{-7} M_{\odot}$ (green circles). The global dipole moments for all of these solutions are less buried than the representative example (Fig. \ref{fig:repe}), with $\tilde{\mu}_{1} / \mu_{\star} \geq 0.48$, since they have less accreted mass. However, in all cases with $M_{a} \gtrsim 10^{-7} M_{\odot}$, the octupole moment is greatest, with $\tilde{\mu}_{3}/ \mu_{\star} = 0.76$ for $M_{a} = 1.2 \times 10^{-7} M_{\odot}$. The prominence of higher-order moments decreases for lower accreted masses, as can be seen for $M_{a} = 8 \times 10^{-8} M_{\odot}$, where even the $\ell = 5$ moment is small compared to its dipole and octupole counterparts (with $\tilde{\mu}_{5}= 0.13 \mu_{\star} \ll \tilde{\mu}_{3}$). Overall, moments with $\ell \geq 9$ are negligible for accreted masses $M_{a} \lesssim 1.2 \times 10^{-7} M_{\odot}$. For the narrow range $8 \times 10^{-8} \leq M_{a} / M_{\odot} \leq 1.2 \times 10^{-7}$ considered in Fig. \ref{fig:symevarymass}, we do not see a systematic trend of $\tilde{\mu}_{\ell}$ versus $M_{a}$ but rather see that the multipolarity is similar in all three cases, as expected. Widening the range of $M_{a}$ substantially, for example to include the regime $M_a \gg M_c$, is impractical due to numerical convergence issues, as described in detail elsewhere \citep{pm04,vigstab08,vig09,pri11}.

\begin{figure}
\includegraphics[width=0.473\textwidth]{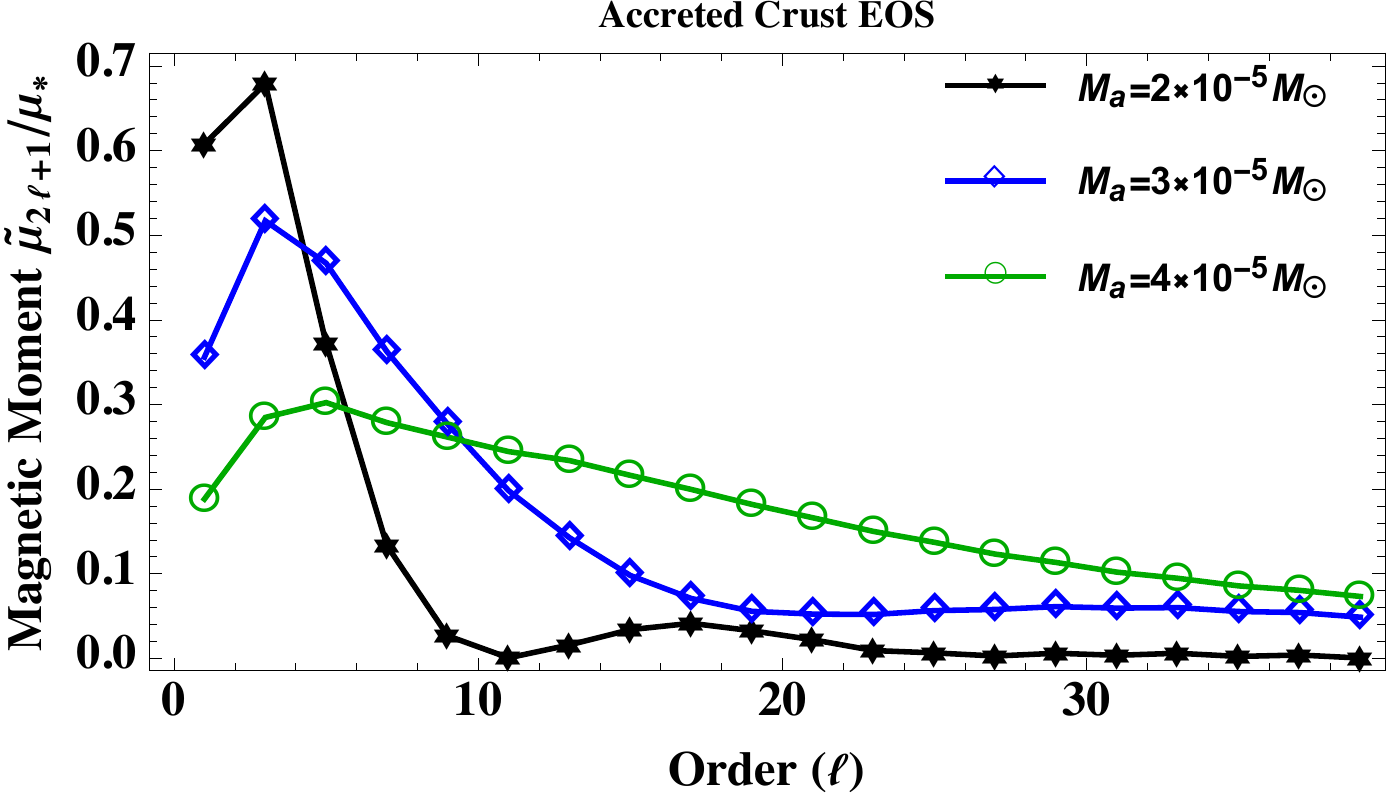}
\caption{Dimensionless multipole moments $\tilde{\mu}_{\ell}/\mu_{\star}$ versus multipole order $\ell$ for $M_{a} = 2 \times 10^{-5} M_{\odot}$ (black stars), $M_{a} = 3 \times 10^{-5} M_{\odot}$ (blue diamonds), and $M_{a} = 4 \times 10^{-5} M_{\odot}$ (green circles) for the accreted crust EOS with $\Gamma = 1.18$ and $K = 6.18 \times 10^{14}$. \label{fig:symacc}
}
\end{figure}

For comparison, Figure \ref{fig:symacc} shows moments for a somewhat wider variety of accreted masses [$M_{a} = 2 \times 10^{-5} M_{\odot}$ (black stars), $M_{a} = 3 \times 10^{-5} M_{\odot}$ (blue diamonds), and $M_{a} = 4 \times 10^{-5} M_{\odot}$ (green circles)] but for the accreted crust EOS built to match the data collated in Table 1 of \cite{haen90}, as detailed in Sec. 2.3. This EOS (with $\Gamma = 1.18$) is softer than the relativistic electron EOS, so that the resulting mountains are more compressed and more mass must accrete to achieve the same degree of burial, with $M_{c} \approx 2.4 \times 10^{-5} M_{\odot}$ \citep{suvmel19}. For $M_{a} \lesssim 3 \times 10^{-5} M_{\odot}$, the octupole moment is dominant (e.g. with $\tilde{\mu}_{3} / \mu_{\star} = 0.52$ for $M_{a} = 3 \times 10^{-5} M_{\odot}$). The $\ell =5$ moment is also comparable to the octupole for $M_{a} \gtrsim 3 \times 10^{-5} M_{\odot}$, and actually exceeds both the $\ell =1 $ and $\ell =3$ moments for the highly-buried case with $M_{a} = 4 \times 10^{-5} M_{\odot}$. In this latter case, the dipole moment is reduced by $\gtrsim 80\%$ and $\tilde{\mu}_{5} / \tilde{\mu}_{3} = 1.06$. The $\ell =7$ moment is also roughly equal to the octupole moment in this case. 

Overall, the moments for this EOS behave in a qualitatively similar way to those seen in Fig. \ref{fig:symevarymass}. However, $M_{a}$ exceeds somewhat the associated characteristic mass $M_{c}$, so higher-order multipoles are non-negligible for a greater range of $\ell$; for the highly-buried $M_{a} = 4 \times 10^{-5} M_{\odot}$ case, the dipole moment is weaker than each of the $\ell \leq 19$ moments. The trend of increasing high-$\ell$ multipole prominence with greater $M_{a}$ is expected physically because field line curvature near the equator increases with $M_{a}$ (see Sec. 5.1). Higher harmonics are required to capture sharp angular gradients, which necessitates the inclusion of non-negligible, higher-$\ell$ terms in the expansion for $\boldsymbol{B}$. Again, extending the simulations further to the regime $M_a \gg M_c$ is impossible due to challenging numerical convergence issues \citep{pm04,vigstab08,vig09,pri11}.

\subsection{Equation of state}

For completeness, we also compare simulations with a fixed accreted mass, though with varying polytropic constants $K$ and $\Gamma$. In particular, at depths where the neutron-drip density $\rho_{\text{nd}} \sim 5 \times 10^{11} \text{ g cm}^{-3}$ is exceeded, non-relativistic, degenerate neutrons are expected to populate the crust in addition to the relativistic, degenerate electrons \citep{brown00}. This leads to a stiffening of the EOS towards the values appropriate for an isentropic gas of degenerate neutrons, viz. $\Gamma = 5/3$ and $K = 1.23 \times 10^{15}$. As such, to represent a crust which contains a mixture of relativistic $(\Gamma = 4/3)$ and non-relativistic $(\Gamma = 5/3)$ neutrons and electrons in some ratio, we consider a range $1.33 \leq \Gamma \leq 1.37$ and $0.49 \leq K / 10^{15} \leq 1.23$.

\begin{figure*}
\includegraphics[width=\textwidth]{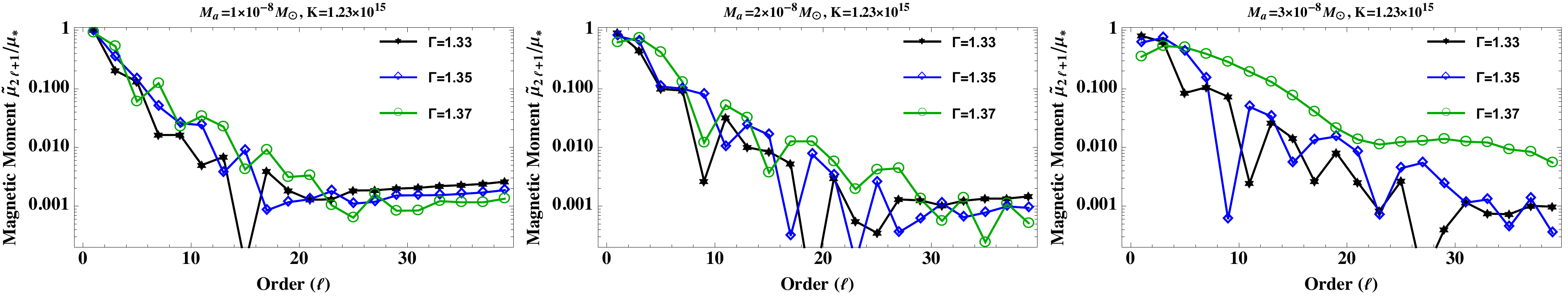}
\caption{Dimensionless multipole moments $\tilde{\mu}_{\ell}/\mu_{\star}$ versus multipole order $\ell$ for a variety of adiabatic indices $1.33 \leq \Gamma \leq 1.37$ (see colour-coded symbols and curves in plot legends) with fixed $K = 1.23 \times 10^{15}$, for accreted masses $M_{a} = 10^{-8} M_{\odot}$ (left panel), $M_{a} = 2 \times 10^{-8} M_{\odot}$ (center panel), and $M_{a} = 3 \times 10^{-8} M_{\odot}$ (right panel). \label{fig:symvarygamma}
}
\end{figure*}

\begin{figure*}
\includegraphics[width=\textwidth]{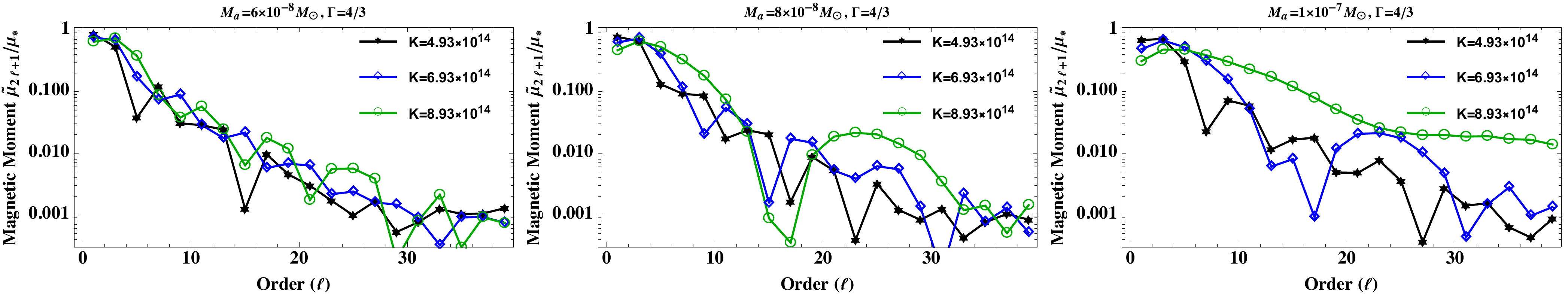}
\caption{Dimensionless multipole moments $\tilde{\mu}_{\ell}/\mu_{\star}$ versus multipole order $\ell$ for a variety of polytropic constants $4.9 \times 10^{14} \leq K \leq 8.9 \times 10^{14}$ (see colour-coded symbols and curves in plot legends) with fixed $\Gamma = 4/3$, for accreted masses $M_{a} = 6 \times 10^{-8} M_{\odot}$ (left panel), $M_{a} = 8 \times 10^{-8} M_{\odot}$ (center panel), and $M_{a} = 10^{-7} M_{\odot}$ (right panel). \label{fig:symvaryk}
}
\end{figure*}

Figure \ref{fig:symvarygamma} details the moments for Grad-Shafranov equilibria with fixed $K = 1.23 \times 10^{15}$ but with $\Gamma = 1.33$ (black stars), $\Gamma = 1.35$ (blue diamonds), and $\Gamma = 1.37$ (green circles), with $M_{a}$ increasing from the left panel to the right. Larger $\Gamma$ means a harder EOS and a less compressed mountain in general, so that polar magnetic flux is squeezed into comparatively larger volumes than for softer EOS. Consequently, screening currents flow further from the stellar surface, and $\mu$ reduces more for a given $M_{a}$ \citep{pri11,suvmel19}. As before, increasing $M_{a}$ shifts the magnetic multipolarity towards higher-orders: $\mu_{1}$ decreases relative to $M_{a}$, and is superseded as the leading-order multipole by the octupole (e.g. $\tilde{\mu}_{3}/ \mu_{\star} = 0.51$ for $M_{a} = 3 \times 10^{-8}$). An exception is the lightly-buried case with $M_{a} = 10^{-8} M_{\odot}$, where even for $\Gamma = 1.37$ the dipole moment is only buried by $10\%$. In this latter case, all $\ell > 3$ moments are negligible, being less than $\sim 0.1$ times the value of the dipole component, indicating that multipolarity can be ignored for low accreted masses regardless of the particulars of the EOS [cf. \cite{singh20}]. In contrast, for accreted masses $M_{a} \gtrsim 3 \times 10^{-8} M_{\odot}$ and sufficiently stiff EOS with $\Gamma \geq 1.37$, moments up to $\ell \sim 9$ are comparable with the buried dipole moment; e.g., $\tilde{\mu}_{9} / \tilde{\mu}_{1} = 0.82$ for the green circles run shown in the rightmost panel.

Figure \ref{fig:symvaryk} is similar to Fig. \ref{fig:symvarygamma}, though we fix $\Gamma = 4/3$ and consider the range $4.9 \leq K / 10^{14} \leq 8.9$, where again $M_{a}$ increases from the left to right panels. These simulations reinforce the conclusions drawn from Figs. \ref{fig:symevarymass} and \ref{fig:symvarygamma}, where we see that the octupole moment dominates over the other moments until $M_{a}$ reaches a sufficiently large, EOS-dependent value, whereupon the $\ell =5$ moment matches the octupole, i.e. $\tilde{\mu}_{5} / \tilde{\mu}_{3} = 0.95$ for $M_{a} = 10^{-7} M_{\odot}$ for the $K =8.9 \times 10^{14}$ case. In particular, the fact that the same patterns are observed within Figs. \ref{fig:symvarygamma} and \ref{fig:symvaryk} indicates that different combinations of $M_{a}$ and the EOS parameters can lead to almost exactly the same set of moments. A larger accreted mass can be mimicked by a stiffer EOS and vice-versa since both enhance field line compression. For instance, fixing $\Gamma = 4/3$, the moments with $M_{a} = 8 \times 10^{-8} M_{\odot}$,  $K = 8.9 \times 10^{14}$, and $\ell \leq 7$ agree to within $15\%$ with the moments for $M_{a} = 10^{-7} M_{\odot}$ and $K = 6.9 \times 10^{14}$. The higher moments are negligible in both cases, with $\tilde{\mu}_{\ell} / \mu_{\star} \lesssim 10^{-2}$ for $\ell > 7$. Mountain predictions for field multipolarity are therefore relatively insensitive to the exact values of $M_{a}$ and the EOS parameters, and a particular moment set $\{\tilde{\mu}_{\ell}\}$ corresponds either to several distinct mountain configurations or none at all, though time-dependent effects may modify the solution space \citep{brown98,litwin01,vig09}.

\section{Asymmetric accretion: non-antipodal hot spots}

As discussed in Sec. 1, the recent NICER results for the recycled pulsar PSR J0030$+$0451 indicate that the system has non-antipodal hot spots \citep{bilous19}. This implies that the magnetic field cannot be that of a simple dipole and, more importantly, that the overall magnetic field cannot be north-south symmetric. In the context of magnetic burial, this requires an equatorially-asymmetric accretion flow and/or cross-hemispheric mass transport\footnote{A third possibility not considered here is that the pre-accretion field is not north-south symmetric. Internal processes [such as crustal Hall drift \citep{gep13}], possibly accelerated by accretion \citep{urp95}, may drive an asymmetric field evolution during an LMXB phase even if the accretion flow is symmetric.}. Such a scenario is considered in this section, where we allow for asymmetric accretion, wherein the mass-flux profile $M(\psi)$ differs between the northern and southern hemispheres. This allows for even multipoles to be non-zero. 

To this end, we consider both the relativistic electron $(\Gamma = 4/3, K = 4.93 \times 10^{14})$ and accreted crust $(\Gamma = 1.18, K = 6.13 \times 10^{14})$ EOS used in Sec. 4.2, though we take different $b$ and $M_{a}$ values between the two hemispheres when calculating $d M / d \psi$ through expression \eqref{eq:massfluxexp}. We stress that the models presented here do not represent a realistic accretion process. They simply demonstrate that the burial scenario accommodates non-antipodal hot spots qualitatively.

\subsection{Representative example}

We first consider a representative example for the accreted crust EOS with $\mas = 2 \man = 2 \times 10^{-5} M_{\odot}$, $\bn = 3$ and $\bs = 4$. Physically speaking, $\bs > \bn$ implies that a smaller fraction of $\mas$ accretes within the southern polar cap $0 \leq \psi(r,\theta \geq \pi /2) \leq \psi_{a}^{\text{S}}$ than $\man$ accretes within the northern cap, though the accretion flow is such that more material by mass is deposited within the southern hemisphere overall, viz. $\mas > \man$. From equation \eqref{eq:bformula}, we see that a value of $\bs = 4$ implies that $\approx 64\%$ of $\mas$ accumulates within the southern polar-flux tube, while $\approx 67\%$ of $\man$ is loaded onto the northern polar-flux tube for $\bn = 3$. As such, the section of the mountain contained within the southern polar cap contains only $\approx 94\%$ more material by mass than its northern counterpart even though we have $\mas = 2 \man$, because we have $ \left( e^{4} - e \right) \mas / \left[ \left( e^{4} - 1 \right) \man \right] = 1.94$.

\begin{figure*}
\includegraphics[width=\textwidth]{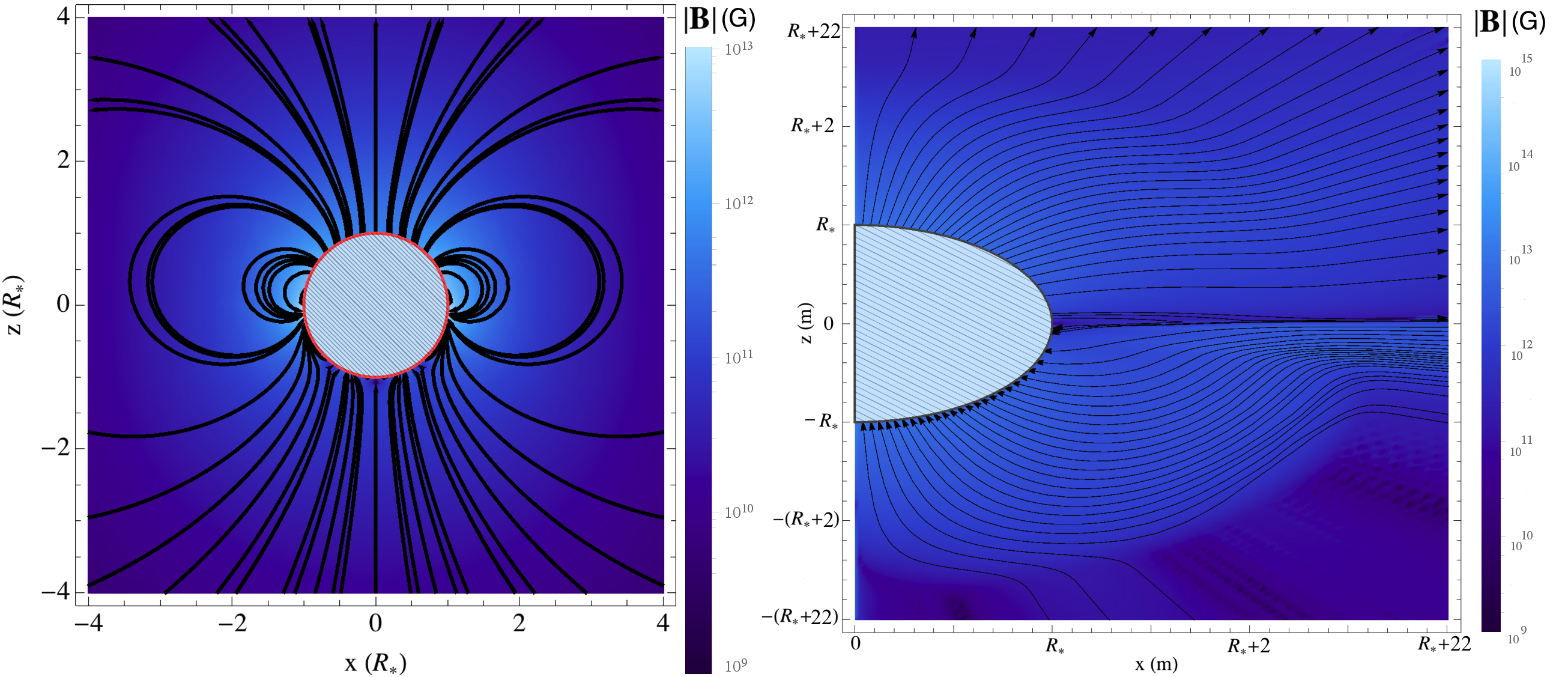}
\caption{Magnetic field lines for the post-accretion field for a mountain with the accreted crust EOS $(\Gamma = 1.18, K = 6.13 \times 10^{14}$) with $M_{a} = 10^{-5} M_{\odot}$ in the north and $M_{a} = 2 \times 10^{-5} M_{\odot}$ in the south, and with $b=3$ in the north and $\bs=4$ in the south. The colour scales show $|\boldsymbol{B}|$, with brighter shades indicating a stronger field. The neutron star surface is shown by the grayed-out surface centered at the origin. Field lines in the mountain region $\rs \leq r \leq \Rm$ (for $x>0$; right panel) are contained within the thin, red layer surrounding the star (left panel). The mountain achieves a maximum radial extent of $\Rm = 22 \text{ m}$ in this case.}
\label{fig:asymb}
\end{figure*}

Figure \ref{fig:asymb} shows field lines (solid lines) in a meridional cross-section for the force-free sum of multipoles (up to order $\ell = 51$; left panel) fitted to the Grad-Shafranov output $\psi$ (right panel) at the computational boundary $\rf$, where the colour scale shows the strength $|\boldsymbol{B}|$ of the post-burial field \eqref{eq:magfield}. Similar to Fig. \ref{fig:symb}, the neutron star itself is represented by the grayed-out surface centered at the origin, while the mountain occupies the thin, red region surrounding the star in the left panel. The mountain achieves a height of $\Rm = 22 \text{ m}$ for this particular simulation, roughly $10\%$ of that of the mountain in Fig. \ref{fig:symb} even though the accreted mass is two orders of magnitude larger, because the EOS here is softer. 

\begin{figure}
\includegraphics[width=0.497\textwidth]{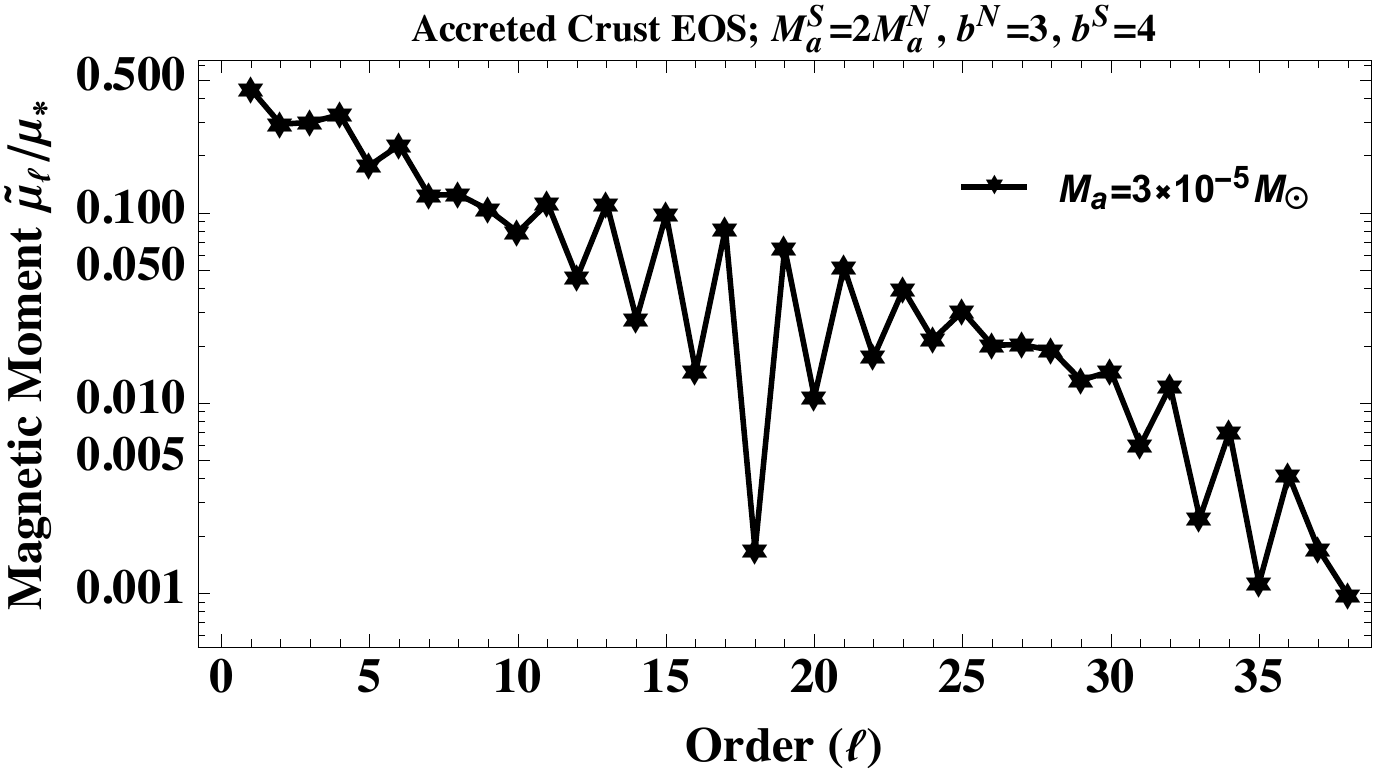}
\caption{Dimensionless multipole moments $\tilde{\mu}_{\ell}/\mu_{\star}$ versus multipole order $\ell$ for an asymmetric accretion model with $M_{a} =10^{-5} M_{\odot}$ and $b=3$ in the north and $M_{a} = 2 \times 10^{-5} M_{\odot}$ and $b=4$ in the south. The analogous graph for symmetric accretion is presented in Fig. \ref{fig:symacc}. }
\label{fig:asymmult}
\end{figure}

Several features unique to asymmetric accretion are evident in Fig. \ref{fig:asymb}. The dipolar loops surrounding the equator are tilted northward, as the buried field in the south ($z<0$) squashes the magnetic field into the northern hemisphere, overwhelming the total pressure (thermal plus magnetic) at the base of the lighter, northern mountain. The emergence of even-order multipoles (especially the quadrupole, with $\tilde{\mu}_{2} / \mu_{\star} = 0.4$; see Fig. \ref{fig:asymmult}) is evident, as field line compression in the southern hemisphere is much greater than in the northern hemisphere, especially near the (geographic) equator, which breaks the equatorial reflection symmetry. The field in the mountain region resembles a lop-sided magnetic tutu, and the curvature of field lines anchored near the southern (geographic) pole is substantial, as $\boldsymbol{B}$ exhibits strong gradients there (reaching a maximum $|\boldsymbol{B}| \sim 10^{15}$ G within the mountain region). In the exterior region $r \geq \Rm$, the maximum field strength is $\sim 5$ times higher in the southern hemisphere (maximum $|\boldsymbol{B}| = 1.1 \times 10^{13} \text{ G}$ at $\theta = 1.69$) than in the northern hemisphere (maximum $|\boldsymbol{B}| = 2.9 \times 10^{12} \text{ G}$ at $\theta = 1.04$).

Figure \ref{fig:asymmult} shows the normalised moments $\tilde{\mu}_{\ell}$ for the field displayed in Fig. \ref{fig:asymb}. Even moments are non-zero in this case. The quadrupole, octupole, and $\ell=4$ moments take values within a few percent of each other, though are weaker than the dipole component by $\sim 20\%$. Comparing the moments with the $M_{a} = 3 \times 10^{-5} M_{\odot}$ symmetric case (blue diamonds in Fig. \ref{fig:symacc}), we see that the dipole moment is comparatively favoured by the asymmetry, with $\tilde{\mu}^{\text{asym}}_{1} / \tilde{\mu}^{\text{sym}}_{1} = 1.25$. This occurs because of the field line crowding effect seen in Fig. \ref{fig:asymb}. The global dipole moment is sensitive to field line curvature at both poles, so $\bs > \bn$ and $\mas > \man$ imply that lines are combed sideways but tilted northward, and the dipole component is overall less affected than in the symmetric case. In the symmetric case, even for $M_{a} \lesssim 2 \times 10^{-5} M_{\odot}$ where the dipole moment is buried by $\lesssim 40\%$, the octupole moment satisfies $\tilde{\mu}_{3} > \tilde{\mu}_{1}$. In the asymmetric case, even though the dipole moment is buried by $\gtrsim 55\%$, it exceeds the octupole, viz. $\tilde{\mu}_{3} / \tilde{\mu}_{1} = 0.67$. This indicates that the growth of even-order moments comes at the expense of odd-order moments, i.e. the non-negligible ($\tilde{\mu}_{2} \sim 0.3$) quadrupole component detracts from the octupole component. This is likely because field line curvature in the northern hemisphere decreases due to the lop-sided accretion, reducing the odd-moments relative to the symmetric case because closed field lines are less pinched overall.

\begin{figure}
\includegraphics[width=0.497\textwidth]{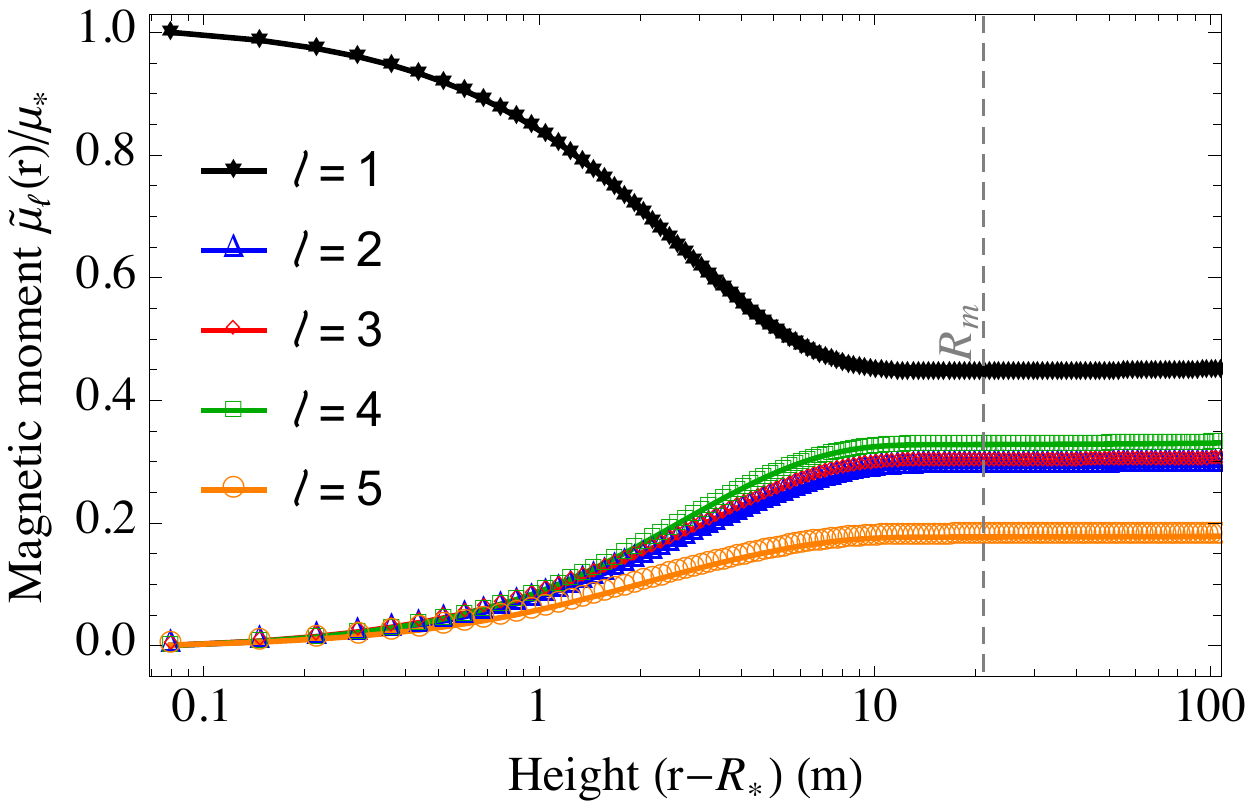}
\caption{The five highest-order moments as functions of radius, $\tilde{\mu}_{\ell}(r)$ [defined by replacing $\mu(\rf)$ by $\mu(r)$ on the right-hand side of \eqref{eq:normmom}], throughout the region $\rs \leq r \leq 5 \Rm$ containing the diamagnetic currents in the mountain for the asymmetric model with the accreted crust EOS, $\mas = 2 \man = 2 \times 10^{-5} M_{\odot}$, $\bn = 3$ and $\bs = 4$. The mountain-atmosphere boundary $\Rm = 22 \text{ m}$ is shown by the dashed, vertical line.}
\label{fig:radialmult}
\end{figure}

Although we focus mostly on the far field in this paper, it is interesting to also monitor the variation of the multipole moments with altitude through the accreted layer, where the diamagnetic screening currents reside (right panel of Fig. \ref{fig:asymb}). Figure \ref{fig:radialmult} shows $\tilde{\mu}_{\ell}$ for $\ell \leq 5$ as a function of $r$ [defined by replacing $\mu(\rf)$ by $\mu(r)$ on the right-hand side of \eqref{eq:normmom}] up to and a bit beyond the mountain-atmosphere boundary, i.e. for $\rs \leq r \leq 5 \Rm$. The higher-$\ell$ moments grow monotonically while the dipole component reduces, until the exterior region at $r \geq \Rm$ is reached where the dipole reaches its buried value $\tilde{\mu}_{1}/ \mu_{\star} = 0.45$. Small variations and oscillations on the order of $\sim 3\%$ occur between $\Rm$ and $\rf$. Comparing with Fig. \ref{fig:asymmult}, where we note that the leading-order non-dipole moment is $\ell =4$, we see in fact that $\tilde{\mu}_{4}$ is the largest non-dipole moment throughout the entire mountain, closely followed by $\tilde{\mu}_{2}$ and $\tilde{\mu}_{3}$ which are within $\lesssim 5\%$ of $\tilde{\mu}_{4}$. We may therefore conclude that the global moment calculations \eqref{eq:multmoms} are not sensitive to the numerical truncation radius $\rf$: fitting a force-free sum of multipoles to $\psi$ at any $r \geq \Rm$ leads to approximately the same moment set.

\subsection{Accreted mass and equation of state}

As in Secs. 4.2 and 4.3, we present here simulations for a variety of accreted masses for the accreted crust and relativistic electron EOS, to further validate the conclusions drawn from the representative example given above.

\begin{figure*}
\includegraphics[width=\textwidth]{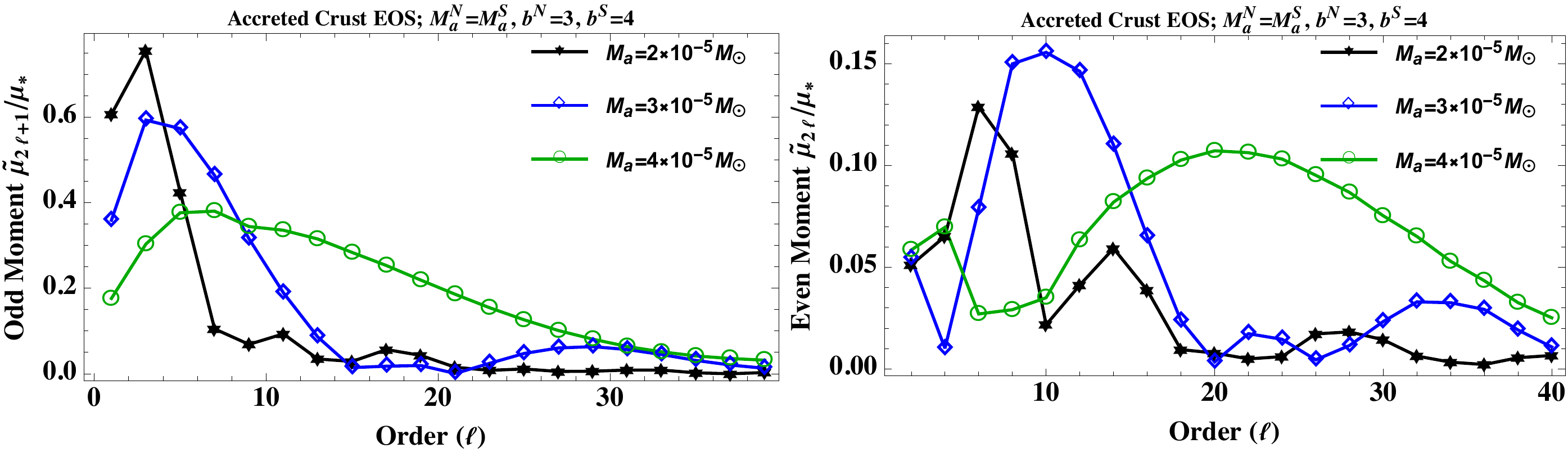}
\caption{Odd- (left panel) and even- (right panel) order dimensionless multipole moments $\tilde{\mu}_{\ell}/\mu_{\star}$ versus multipole order $\ell$ for accreted masses (as in Fig. \ref{fig:symacc})  $M_{a} = 2 \times 10^{-5} M_{\odot}$ (black stars), $M_{a} = 3 \times 10^{-5} M_{\odot}$ (blue diamonds), and $M_{a} = 4 \times 10^{-5} M_{\odot}$ (green circles) for mountains governed by the accreted crust EOS, with hemispheric fluxes \eqref{eq:massfluxexp} with $\bn = 3$ and $\bs = 4$. \label{fig:nonsymac}
}
\end{figure*}

For ease of presentation, Figure \ref{fig:nonsymac} shows odd- (left panel) and even- (right panel) order moments for the accreted crust EOS with $\man = \mas$ but $\bn = 3$ and $\bs = 4$ (similar to Fig. \ref{fig:symacc} though with different $b$ values). Overall, the odd moments are in fact similar to the symmetric case. However field lines from the south encroach into the north (see Fig. \ref{fig:asymb}) and the moments are slightly reduced; we obtain $\tilde{\mu}^{\text{asym}}_{1} / \tilde{\mu}^{\text{sym}}_{1} = 1.07$ for $M_{a} = 4 \times 10^{-5} M_{\odot}$, for example. This occurs because comparatively less material is loaded into the southern polar flux tube for $\bs > \bn$. The even moments are non-zero and increase monotonically with increasing $M_{a}$. They are smaller than neighbouring ($\ell \pm 1$) odd moments. For example, the quadrupole $(\tilde{\mu}_{2} / \mu_{\star} \lesssim 0.06$ for $M_{a} \lesssim 4 \times 10^{-5} M_{\odot})$ is $\gtrsim 5$ times smaller than the octupole. 

\begin{figure*}
\includegraphics[width=\textwidth]{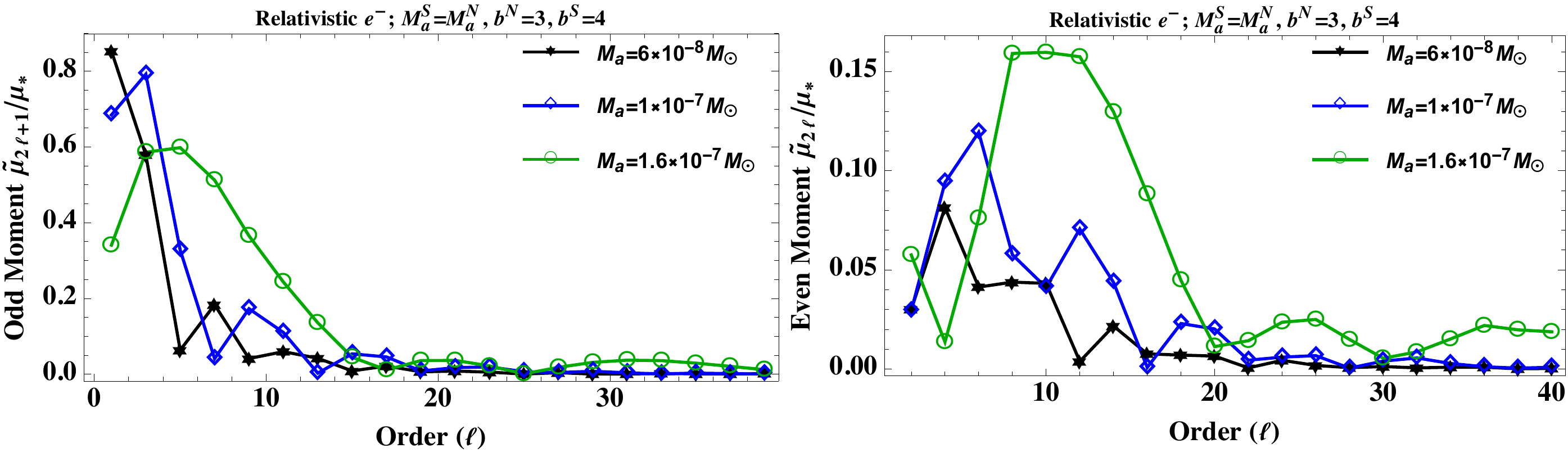}
\caption{As for Fig. \ref{fig:nonsymac} but with a relativistic electron EOS and accreted masses  $M_{a} = 6 \times 10^{-8} M_{\odot}$ (black stars), $M_{a} = 10^{-7} M_{\odot}$ (blue diamonds), and $M_{a} = 1.6 \times 10^{-7} M_{\odot}$ (green squares). \label{fig:nonsyme}
}
\end{figure*}

\begin{figure}
\includegraphics[width=0.493\textwidth]{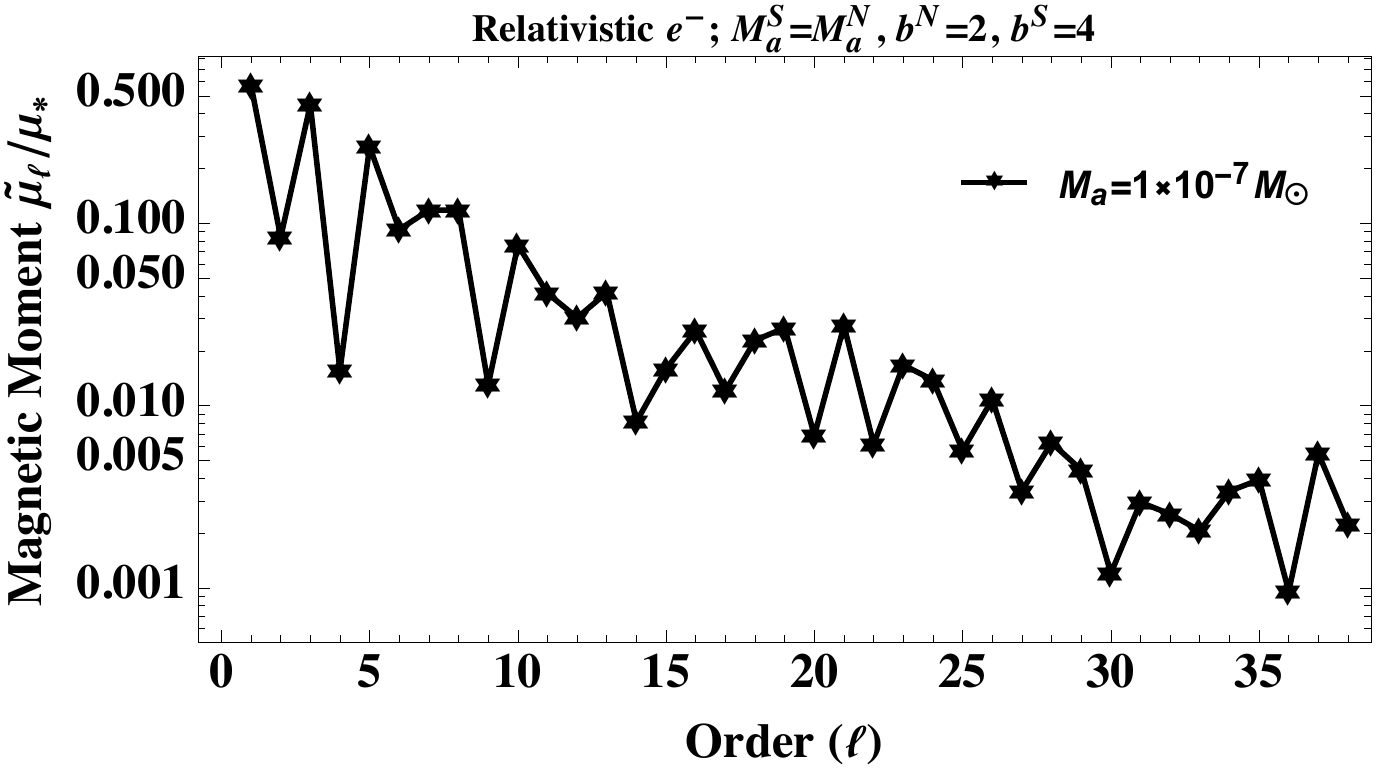}
\caption{Dimensionless multipole moments $\tilde{\mu}_{\ell}/\mu_{\star}$ versus multipole order $\ell$ for a mountain with $M_{a} = 10^{-7} M_{\odot}$ (i.e. $\mas = \man = 5 \times 10^{-8} M_{\odot}$) and the relativistic electron EOS ($\Gamma = 4/3, K = 4.93 \times 10^{14}$) with $\bn = 2$ and $\bs = 4$. \label{fig:nonsymvaryb}
}
\end{figure}

Figure \ref{fig:nonsyme} is similar to Fig. \ref{fig:nonsymac} but employs the relativistic electron EOS, with comparatively lower $M_{a}$ overall so that the dipole moments $\tilde{\mu}_{1}$ are comparable to those shown in Fig. \ref{fig:nonsymac}. The odd moments match with the corresponding symmetric cases described in Sec. 4 to within $\sim 10\%$, and the even-moments, while non-zero, are relatively weak compared to the neighbouring $(\ell \pm 1)$ odd moments, e.g. $\tilde{\mu}_{4} / \tilde{\mu}_{5} = 0.29$ for $M_{a} = 10^{-7} M_{\odot}$. Putting everything together, the burial scenario therefore predicts that a factor $\sim 2$ mass loading differential between the north and south (i.e. $\mas \gtrsim 2 \man$ or vice-versa) is necessary to develop quadrupole moments satisfying $\tilde{\mu}_{2}/ \tilde{\mu}_{1} \sim 1$ independently of the EOS. This conclusion is further validated by Figure \ref{fig:nonsymvaryb}, which depicts a case where the $b$-ratio is $2$ between the north and south rather than $4/3$ as used in previous simulations. Here, the quadrupole component $\tilde{\mu}_{2} / \mu_{\star} = 0.09$ is a factor $\sim 5$ smaller than the octupole and the dipole moment is roughly equal (within a few percent) to the dipole moment for the symmetric $\bn=\bs=3$ case. 

\section{Discussion}

Recycled pulsars may possess magnetically supported polar mountains from prior episodes of accretion and have buried magnetic fields \citep{blon86,pm04,pat12,dip13}. Recent results from the NICER experiment indicate that recycled pulsars have non-antipodal hot spot structures \citep{bilous19} and therefore have non-dipolar magnetic fields \citep{stur71,mus03}, a conclusion supported by radio pulse modelling and Stokes tomography \citep{chung11,burn14}. In this paper we point out that, in the burial scenario, field lines are laterally shifted via flux-freezing during periods of active accretion, which facilitates the growth of multipole moments $\mu_{\ell}$ with $\ell > 1$, which peak at $\ell$ values that depend on $M_{a}$ and the particulars of the crustal EOS (cf. Figs \ref{fig:symvarygamma} and \ref{fig:symvaryk}). If the polar-cap geometries are asymmetric ($\bn \neq \bs$), or if $M_{a}$ is larger in one hemisphere than the other ($\man \neq \mas$), the magnetic field in one hemisphere crowds into the other hemisphere, dragged by the mass sliding sideways under its own weight. The result is an equatorial asymmetry in the magnetic field and a distinctive	 trend in $\tilde{\mu}_{\ell}$ versus $\ell$, as observed in Figs. \ref{fig:asymb} and \ref{fig:asymmult}.

We extend previous studies by implementing boundary conditions (detailed in Sec. 2.4) that allow the mass-flux relation $M(\psi)$ to differ in the northern and southern hemispheres (see Fig. \ref{fig:psifit}). We consider a wide range $(10^{-8} \lesssim M_{a} / M_{\odot} \lesssim 10^{-4})$ of accreted masses and polytropic EOS $p = K \rho^{\Gamma}$ with $0.49 \leq K / 10^{15} \leq 1.23$ (in cgs units) and $1.18 \leq \Gamma \leq 1.37$. We find that for $\man/ \mas \gtrsim 2$ (or equivalently $\mas / \man \gtrsim 2$) a quadrupole moment can be generated which is comparable with the (buried) global dipole moment, $\tilde{\mu}_{2} \sim \tilde{\mu}_{1}$. For example, for the accreted crust EOS, obtained by fitting a polytrope to the simulations of \cite{haen90}, with $M_{a} = 3 \times 10^{-5} M_{\odot}$ but $\man = 2 \mas$, we obtain $\tilde{\mu}_{2} / \tilde{\mu}_{1} = 0.7$ (see Fig. \ref{fig:asymmult}). For this simulation, the octupole and $\ell =4$ moments are within $\sim 20\%$ of the dipole moment and are even more prominent than the quadrupole. For greater accreted masses leading to more burial, the $2 \leq \ell \lesssim 7$ moments come to dominate over the dipole component. For example, for $M_{a} = 4 \times 10^{-5} M_{\odot}$, we find $\tilde{\mu}_{5} \approx 2 \tilde{\mu}_{1}$ (see Fig. \ref{fig:nonsymac}).

The above results are a first step in constructing hot spot models for recycled pulsars or actively accreting LMXB systems undergoing magnetic burial. They show how a set of multipoles $\{ \mu_{\ell} \}$ can be tied to the accretion history of the object,  although the mapping is not one-to-one as discussed at the end of Sec. 4.3. By fitting a force-free magnetosphere model to NICER data for PSR J0030$+$0451, \cite{chen20} found that a mixed, offset dipole-quadrupole field is sufficient to replicate the observed light curves and antipodal structures \citep{bilous19}. This fits well within the burial scenario for $M_{a} \sim M_{c}$, where the dipole component is reduced by a factor $\sim 2$ and multipoles with $\ell \leq 4$ are appreciable, as for the  $\man = 2 \mas$ case discussed above. For accreted masses $M_{a} \gg M_{c}$, the dipole component is significantly reduced and higher multipoles dictate the field structure. For example, the $M_{a} = 1.6 \times 10^{-7} M_{\odot}$ asymmetric accretion simulation shown in Fig. \ref{fig:nonsyme} produces $\tilde{\mu}_{\ell} \gg \tilde{\mu}_{1}$ for $3 \leq \ell \leq 7$. This suggests that the burial scenario can accommodate a wide range of hot spot geometries, including those observed for PSR J0030$+$0451. {Using Monte Carlo methods, \cite{kalap20} found that a diverse range of multipolar geometries can reproduce the observed X-ray light curves for PSR J0030$+$0451, further strengthening this conclusion.}

Different combinations of $M_{a}$, $K$, and $\Gamma$ can produce quantitatively similar sets of multipole moments $\{ \mu_{\ell} \}$. The simulations shown in Figs. \ref{fig:nonsymac} and \ref{fig:nonsyme} are one such example, with $M_{a} = 3 \times 10^{-5} M_{\odot}$ and an accreted crust EOS versus $M_{a} = 1.6 \times 10^{-7} M_{\odot}$ and a relativistic electron EOS respectively. This implies that multiple mountain configurations could theoretically explain the hot spots of PSR J0030$+$0451, and a fine-tuning between parameters is not required. However, it may be difficult therefore to disentangle the EOS parameters from $M_{a}$ and the particulars of the asymmetric accretion history from measurements of $\mu_{\ell}$ for a few $\ell$ values alone. 

The time-varying mass quadrupole of the rotating mountain emits gravitational waves with a characteristic strain \citep{riles13,lasky15}
\begin{equation} \label{eq:gwstrain}
h_{0} \approx 10^{-26} \left( \frac {\epsilon} {10^{-8}} \right) \left( \frac {P} {2 \text{ ms}} \right)^{-2} \left( \frac {d} {1 \text{ kpc}} \right)^{-1}, 
\end{equation}
for a star with spin period $P$. The exact value of $\epsilon$ depends sensitively on the mountain EOS and can vary by several orders of magnitude \citep{pri11,dip12}. In an idealised model, it can be approximated as \citep{mp05}
\begin{equation} \label{eq:epsilon}
\epsilon \approx \frac {5 M_{a}} {4 M_{\star} \left( 1 + 9 M_{a} / 8 M_{c} \right)},
\end{equation}
for stellar mass $M_{\star}$. Upper limits on $h_{0}$ from the Laser Interferometer Gravitational-wave Observatory (LIGO) set $\epsilon \lesssim 4.6 \times 10^{-8}$ for PSR J0030$+$0451 \citep{aasi17}, which is broadly comparable to the ellipiticites \eqref{eq:epsilon} one gets from first principles modelling for soft EOS \citep{mp05,vigstab08,wette10,dip13}. \cite{woan18} have provided evidence that all millisecond pulsars have a minimum ellipticity $\epsilon \gtrsim 10^{-9}$, which may be connected with a permanent mountain having formed during a prior LMXB phase \citep{mast12}. Moreover, asymmetric accretion leads to uneven thermonuclear activity within the crust, thereby modulating cooling rates \citep{wij13,hask15,pons19} and contributes a `thermal' ellipticity to \eqref{eq:gwstrain} from density deformations triggered by compositional gradients \citep{ush00,singh20,osb20}. Current and upcoming gravitational-wave experiments will therefore help to disentangle the effects of $M_{a}$, $K$, and $\Gamma$ on $\mu_{\ell}$, although not necessarily uniquely, and enable further tests of the mountain scenario.

{Finally, it is important to point out that we have only investigated static mountains in this paper. Time-dependent effects related to the accretion flow [e.g., the onset of magnetic turbulence \citep{lasota01,don07,kong11}] and the magnetohydrodynamic equilibration of the mountain itself [e.g., the onset of ballooning or Parker instabilities \citep{litwin01,vig09,dip13}] may adjust the conclusions reached here. At least with respect to the former, the large values of the magnetization parameter $\sigma$ (Fig. \ref{fig:plasmanumbers}) suggest that a force-free approximation for the magnetosphere is reasonable \citep{michel91}. Further investigation on this point would be worthwhile, especially for near-Eddington accretion rates. Modelling dynamic magnetic fields and the evolution of the (force-free) plasma simultaneously would allow one to study both the accretion process and the formation of hot spots self-consistently.}

\section*{Acknowledgements}
We thank Maxim Priymak and Donald Payne for providing access to the Grad-Shafranov solver. {We are grateful to the anonymous referee for providing helpful feedback.} This work was supported by the Alexander von Humboldt Foundation and the Australian Research Council through its Centre of Excellence program (grant number CE170100004) and Discovery Project DP170103625.
\\
\\
Data availability statement: Observational data used in this paper are publicly available in the cited works. The data generated from computations are reported in the body of the paper, and any additional data will be made available upon reasonable request to the corresponding author.

\bsp \label{lastpage}

\end{document}